\documentclass[10pt]{book}

\usepackage{sussp2004}
\usepackage{graphicx}
\usepackage{latexsym}
\usepackage{makeidx}

\begin{document}

\setcounter{section}{1}
\setcounter{page}{1}

\headings{Deeply virtual Compton scattering: results \& future
{$^\diamond$}}
{Deeply virtual Compton scattering: results \& future}
{Wolf-Dieter Nowak}{DESY, D-15738 Zeuthen, Germany}

\vspace{-5mm}
{\bf Abstract.} Access to Generalised Parton Distributions (GPDs) through 
Deeply Virtual Compton Scattering (DVCS) is briefly described. 
Presently available experimental results on DVCS are summarized in 
conjunction with plans for future measurements.\\
\vspace{-5mm}
%
%
\section{Introduction}
%
For more than two decades the momentum and spin composition of the nucleon 
and other hadrons has been investigated by now, preferentially using charged 
leptons as probes. A great variety of measurements was performed in order to 
study the underlying structure of quarks and gluons that constitute the 
fundamental degrees of freedom in Quantum Chromodynamics (QCD). Their 
momenta and angular momenta cannot yet be calculated from 
first principles, they are encoded in universal distribution 
functions whose determination is a central topic that embraces particle 
physics and hadron physics. 

The longitudinal momenta and polarisations carried by quarks, antiquarks 
and gluons within a fast moving hadron are encoded in universal Parton 
Distribution Functions (PDFs). They are conveniently 
introduced in the description of the inclusive deep inelastic 
scattering (DIS) process, $e \, p \, \rightarrow \, e \, X$. The exchange
of one virtual photon dominates this reaction at fixed-target 
kinematics with center-of-mass energies $\sqrt{s} = \cal{O}$(10~GeV),
and it is still the major contribution at collider kinematics with
$\sqrt{s} = \cal{O}$(300~GeV). In the Bjorken limit of high virtuality 
$Q^2$ and large energy $\nu$ of the photon in the target rest frame, at a 
finite ratio $x_B=\frac{Q^2}{2m \, \nu}$ (m being the nucleon mass), 
the cross section factorises into that of a hard partonic
subprocess and (a certain combination of) PDFs. For a parton of a given 
species, the unpolarised PDF represents the probability of finding it at 
a given fraction $x_B$ of the nucleon momentum, while the polarised one
describes the imbalance of probabilities between oppositely polarised 
partons. PDFs are called `forward' distributions, as the
inclusive $\gamma^* \, p$ cross section can be expressed through the
optical theorem by the imaginary part of the forward Compton amplitude 
$\gamma^* \, p \rightarrow \, \gamma^* \, p$.

{\footnotesize $^\diamond)$Invited talk at Hadron Physics I3 Topical Workshop,
Aug.30-Sept.1, 2004, St Andrews, Scotland}
%
%
\section{Generalised Parton Distributions}
\label{sec:GPDs}
%
%
%
The theoretical framework of Generalised Parton Distributions
(Dittes \etal 1988; M\"uller \etal 1994;  Radyushkin 1996; Ji 1997; 
Bl\"umlein \etal 1999) 
is capable of simultaneously treating several types of processes ranging 
from inclusive to hard exclusive lepton-nucleon scattering. Exclusive 
scattering is `non-forward' in nature since the photon initiating the 
process is virtual and the final-state particle is usually real, forcing 
a small but finite $t$, the squared momentum transfer between initial 
and final nucleon states. GPDs depend on $t$ and on $Q^2$, the hard 
scale of the process, and also on two longitudinal momentum variables. 
Through these dependences they carry information on two-parton correlations 
and on quark transverse spatial distributions (Burkardt 2000; Ralston 
and Pire 2002; Diehl 2002; Belitsky and M\"uller 2002, Burkardt 2003).
A recent comprehensive theoretical review can be found in 
(Diehl 2003).

Presently the most intensely discussed GPDs are the chirally-even, 
or quark-helicity conserving GPDs $F^q$ ($F=H, \tilde{H}, E, 
\tilde{E}$ and $q=u,d$). In order to constrain their non-forward 
behaviour, measurements can be performed of hard exclusive leptoproduction 
of a photon or meson, in processes leaving the target intact. The production 
of a real photon, {\em ie}, Deeply Virtual Compton Scattering (DVCS) 
$e \, p \, \rightarrow \, e \, p \, \gamma$, has two benefits: \\
i) it is considered to be the theoretically cleanest process that can 
be accessed experimentally in the foreseeable future and \\
ii) effects of next-to-leading order (Belitsky and 
M\"uller 1998; Ji and Osborne 1998; Mankiewicz \etal 1998) and sub-leading 
twist (see {\em eg} Anikin \etal 2000; Radyushkin and Weiss 2000; 
Belitsky \etal 2002) are under theoretical control. 

In the generalised Bjorken limit of large photon virtuality $Q^2$ at fixed
$x_B$ and $t$ the dominant pQCD subprocess of DVCS is described by the 
`handbag' diagram shown in the left panel of 
\begin{figure}[htb]
\label{fig:handbag}
\includegraphics[width=9.5cm]{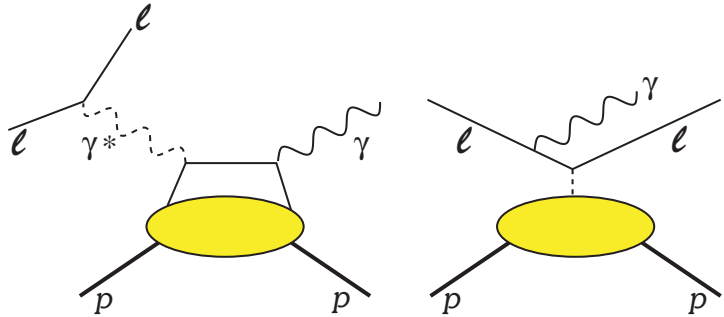}
\centering
\caption{\it Left: Deeply Virtual Compton Scattering. Right: 
Bethe-Heitler process (photon radiated by incoming or outgoing lepton).}
\end{figure}
\begin{figure}[htb]
\vspace*{0.2cm}
\hspace{3.3cm}
\label{fig:GPDwheel}
\includegraphics[width=11.5cm]{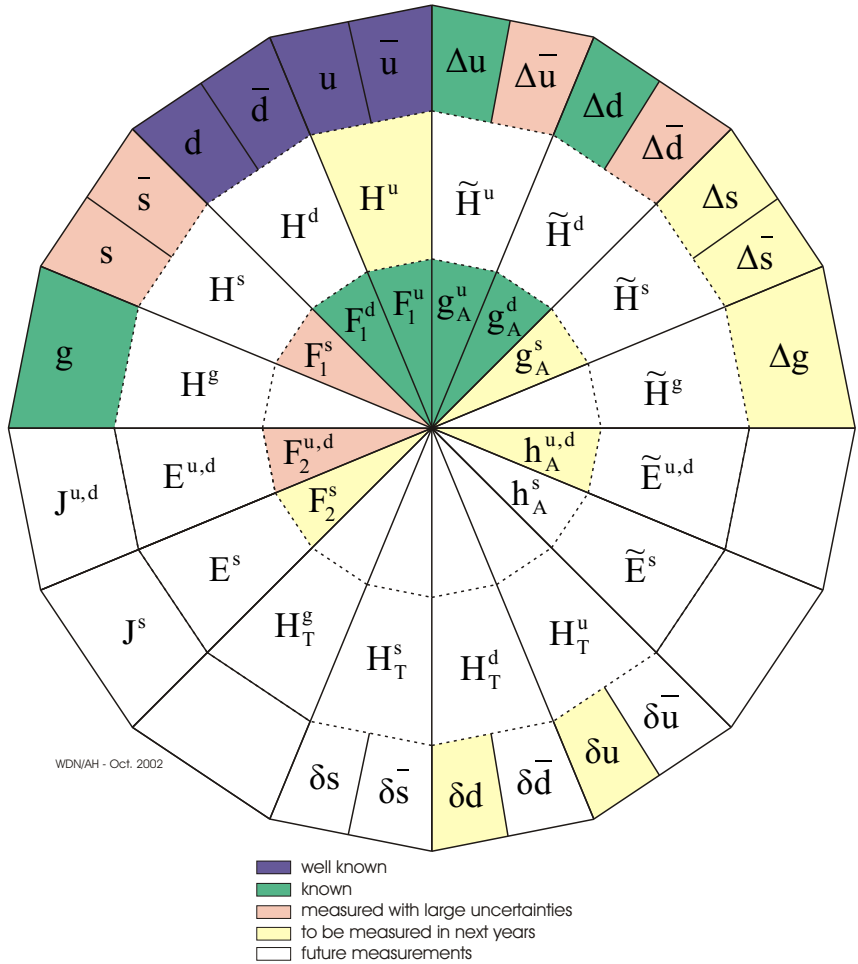}
\centering
\caption{\it Visualisation of (most of) the relevant Generalised Parton 
             Distributions and their limiting cases, forward Parton
             Distributions and Nucleon Form Factors. Different colours
             illustrate the status of their experimental access 
             (see legend). For explanations see text. The Figure has 
             been taken from (Nowak 2003).}
\end{figure}
%
Figure~1. The internal variable $x$ and the skewedness parameter $\xi$, 
with $\xi \simeq \frac{x_B}{2-x_B}$ in the Bjorken limit, describe the 
longitudinal momentum transfer between two partons: the parton (of flavour 
$q$) taken out of the proton carries the longitudinal momentum fraction 
$x+\xi$ and the one put back into the proton carries the fraction $x-\xi$. 
The GPD $F^q(x,\xi,t,Q^2)$ can then be considered as describing the 
correlation between these two partons at the given values of $t$ and $Q^2$.

GPDs reduce to ordinary PDFs in the forward limit, {\em ie}, at
vanishing momentum transfer. The first $x$-moments 
of GPDs are related to certain form factors measured in elastic 
lepton-nucleon scattering which describe the difference of the 
electromagnetic nucleon structure from that of a point-like spin-1/2 
particle. A particular second moment of GPDs, for a given parton species 
$f=(u,d,g)$, is in the limit of vanishing $t$ connected to the total 
angular momentum carried by these partons (see Equation~\ref{eq:JiSumRule}). 
The latter finding (Ji 1997) stimulated strong interest in GPDs, as the total 
angular momenta carried by quarks and gluons in the nucleon
constitute the hitherto missing pieces in the puzzle representing the 
momentum and spin structure of the nucleon. 

Generalised Parton Distributions, as phenomenological functions, have to 
be parameterised. Two ans\"atze are most customary at present:

i) originally, the `factorised ansatz' uses uncorrelated dependences on 
$t$ and $(x,\xi)$.
The former is written in accordance with proton elastic form factors and 
the latter is based on double distributions (Radyushkin 1999) plus additional
D-term (Polyakov and Weiss 1999). Double distributions are 
constructed from ordinary PDFs complemented with a profile function that 
characterises the strength of the $\xi$-dependence; in the limit 
$b \rightarrow \infty$ of the profile parameter $b$ the GPD is 
independent on $\xi$. Note that $b$ is a free parameter to be determined by
experiment, separately for valence and sea quarks.

ii) measurements of elastic diffractive processes and, more recently, 
phenomenological considerations  (Diehl \etal 2005; Guidal \etal 2004)
suggest that the $t$-dependence of the $\gamma^* p$ cross section is
entangled with its $x_B$-dependence. The `Regge ansatz' for GPDs hence uses
for the $t$-dependence of double distributions a soft Regge-type 
parameterisation $\sim |x|^{-\alpha ' \, t}$ with 
$\alpha_{soft} ' = 0.8 ... 0.9$~GeV$^2$ for quarks.

The scheme presented in 
Figure~1 
visualises the present experimental knowledge 
on the above mentioned functions. As main ingredients, GPDs are placed in the 
middle of three concentric rings. Their forward limits and moments
are situated in the adjacent rings: PDFs in the outermost and nucleon 
form factors in the innermost one. Today's experimental 
knowledge of the different functions is illustrated in different colours
from light (no data exist) to dark (well known). The emphasis in 
Figure~1
is placed on the physics message and not on completeness; some GPDs have been 
omitted. Empty sectors mean that the function does not exist, 
decouples from observables in the forward limit, or no strategy is known for
its measurement. More details can be found in (Nowak 2003).
%
%
%
%
\section{Deeply Virtual Compton Scattering}
\label{sec:DVCS}
%
\subsection{Compton Form Factors}
\label{subsec:CFFs}
%
%
%
The Bethe-Heitler (BH) process, or radiative elastic scattering, is 
illustrated in the right panel of Figure~1.
Its final state is indistinguishable from that of the DVCS process, 
hence both mechanisms have to be added on the amplitude level. The 
differential real-photon leptoproduction cross section is given as
\begin{eqnarray}
\label{eq:DVCS-BH-amplitude}
\frac{d\sigma}{dx_BdQ^2d|t|d\phi} & 
\propto & |\tau^{}_{BH}|^2 + |\tau^{}_{DVCS}|^2 + 
\underbrace{\tau^{}_{DVCS} \tau^{*}_{BH}+\tau^{*}_{DVCS} \tau^{}_{BH}}_{I}.
\end{eqnarray}
Here $\phi$ is the azimuthal angle between the scattering plane, 
spanned by the incoming and outgoing leptons, and the
production plane spanned by the virtual photon and the produced real 
photon (cf. 
Figure~3). 
The BH amplitude $\tau^{}_{BH}$ is exactly calculable using the knowledge 
of the elastic nucleon form factors. The DVCS contribution 
$|\tau^{}_{DVCS}|^2$ can then be extracted by integrating over the 
azimuthal dependence of the cross section. In this case the 
interference term $I$ vanishes to leading order in $1/Q$; 
its total contribution at collider kinematics was estimated to be
at the percent level (Belitsky \etal 2002).
\begin{figure}[htb]
\label{fig:kinPlane}
\includegraphics[width=7cm]{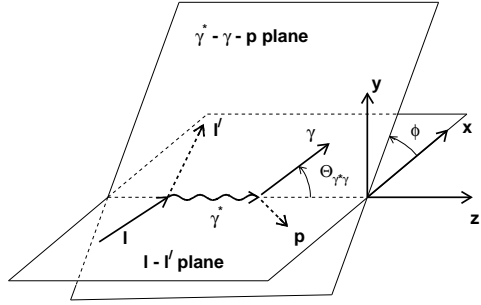}
\centering
\caption{\it Definition of the azimuthal angle $\phi$ in DVCS 
in the target rest frame.}
\end{figure}

The twist-2 DVCS amplitudes can be represented in the convention of 
(Belitsky \etal 2002) as linear combinations of $F_1$ and $F_2$, the 
Dirac and Pauli elastic nucleon form factors, with the Compton form factors 
(CFFs)
$\mathcal{H},\mathcal{E},\widetilde{\mathcal{H}},\widetilde{\mathcal{E}}$
(cf. Equations~\ref{eq:Mtilde},\ref{eq:LTSA3},\ref{eq:TTSA_N},\ref{eq:TTSA_S}).
These complex CFFs are flavour sums of convolutions of the corresponding 
leading-twist GPDs with the hard scattering kernels $C_q^{\mp}$ 
that are available up to NLO in pQCD  (Belitsky and M\"uller 1998; Ji and 
Osborne 1998; Mankiewicz \etal 1998):
\begin{eqnarray}
\label{eq:CFF}
\mathcal{F}(\xi,t,Q^2) = \sum_{q} \int_{-1}^1 dx \; 
                             C^\mp_q(\xi,x)\, F^q(x,\xi,t,Q^2).
\end{eqnarray}
Here the $-(+)$ sign in the superscript applies to the CFFs
$\mathcal{F}=\mathcal{H},\mathcal{E} \;
(\widetilde{\mathcal{H}},\widetilde{\mathcal{E}})$, corresponding to the GPDs 
$F^q=H^q,E^q \; (\widetilde{H^q},\widetilde{E^q})$. 

The real and imaginary parts of a CFF
have different relationships to (the flavour sum over) the respective quark 
GPDs which are embodied. Taking Equation~\ref{eq:CFF} at leading order in 
$\alpha_s$ (Belitsky \etal 2002), the imaginary part
\begin{equation}
\label{eq:CFF1a} 
\mathrm{Im} \left\{ \mathcal{F} \right\} =  \pi 
\sum_{q} e_q^2\left( {F^q}(\xi,\xi,t,Q^2) \mp {F^q}(-\xi,\xi,t,Q^2)\right)
\end{equation}
{\em directly} probes the respective GPDs along the line $x=\pm\xi$.
In contrast, through the real part of the CFF,
\begin{eqnarray}
  \label{eq:CFF2a}
  \mathrm{Re} \left\{ \mathcal{F} \right\} = 
  - \sum_{q} e_q^2 \left[ P \int_{-1}^1 dx \ {F^q}(x,\xi,t,Q^2) 
    \left( \frac{1}{x-\xi} \pm \frac{1}{x+\xi} \right) \right],
\end{eqnarray}
the integral over the respective GPDs is accessed, whereby the weighting 
by the propagators $1/(x\mp\xi)$ strongly enhances the contribution close to
the line $x=\pm\xi$. The sign convention is the same as for
Equation~\ref{eq:CFF} and $P$ denotes Cauchy's principal value. 

Equations~\ref{eq:CFF1a} and \ref{eq:CFF2a} show that in DVCS a GPD, at
given values of $t$ and $Q^2$, is essentially probed along the line 
$x=\pm\xi$, {\em ie,} a complete mapping of a GPD in the 
$(x, \xi)$-plane is impossible and models of GPDs are to be 
constructed to calculate observables that have to be compared to 
corresponding experimental results in an iterative procedure. 

\noindent Full $(x,\xi)$-mapping of GPDs is still possible, 
at least in principle:

\noindent i) once a large enough dynamic range in $Q^2$ is available in
DVCS measurements, the known $Q^2$-evolution of GPDs can be used to constrain 
their $x$-dependence, similar as for the extraction of ordinary PDFs in DIS.

\noindent ii) in hard exclusive leptoproduction of a {\em virtual} photon
(double DVCS or DDVCS) its virtuality, {\em ie} the effective 
mass of the produced lepton pair, is an additional variable that 
facilitates a complete mapping of GPDs. However, the DDVCS cross section 
is suppressed by an additional factor $\alpha_{em}^2$, thereby
making this reaction practically inaccessible in the foreseeable future
(Guidal 2002).

The $t$-dependence of GPDs is directly accessible in DVCS
although high experimental precision, {\em ie}, high statistical accuracy in 
conjunction with sufficient resolution is required to extrapolate to the 
limes $t \rightarrow 0$.
The latter is of particular importance for the evaluation of the 2nd 
$x$-moment of the two `unpolarised' GPDs $H^f+E^f$, which is related to the 
total angular momentum $J^f$ of the parton species $f=(u,d,g)$, at a given 
value of $Q^2$ (Ji 1997):
\begin{eqnarray}
\label{eq:JiSumRule}
J^f(Q^2) = \lim_{t \rightarrow 0} \frac{1}{2} 
  \int_{-1}^{1} dx \ x \left[H^f(x,\xi,t,Q^2) + E^f(x,\xi,t,Q^2)\right].
\end{eqnarray}
%
%
%
%
\subsection{The Interference Term}
\label{subsec:InterferenceTerm}
%
%
%
The interference term $I$ is of special interest, as the measurement of
its azimuthal dependence opens experimental access to the {\em complex} 
DVCS amplitudes, {\em ie}, to {\em both} their magnitude and phase 
(Diehl \etal 1997). 
This method of using the BH process as an `amplifier' to study DVCS
can be compared to holography (Belitsky and M\"uller 2002) in the sense that 
the phase of the Compton amplitude is measured against the known `reference
phase' of the BH process. 

The full interference term can be filtered out by
forming a cross section asymmetry, or difference, w.r.t. the charge of the 
lepton beam (Brodsky \etal 1972). The imaginary part of the interference
term can be accessed by forming 
single-spin asymmetries, or differences, w.r.t. the spin of the lepton beam
(Kroll \etal 1996) or of the target (Belitsky \etal 2002; Diehl 2003). 
Note that the measurement of cross section differences is favoured by 
theorists over that of asymmetries (Diehl 2003). Differences are free 
from azimuthal dependences of BH, DVCS and interference terms appearing 
in the denominator of an asymmetry and thereby complicating the separation 
of the relevant terms in the numerator. They allow easier 
separation of higher harmonics when compared to the evaluation of an 
asymmetry, while larger experimental systematic uncertainties may appear. 

Each of the three terms in Equation~\ref{eq:DVCS-BH-amplitude} can be 
expressed as a Fourier series in $\phi$ (Diehl \etal 1997; Belitsky 
\etal 2002). For an unpolarised target, the interference term $I$ can be 
written as
\begin{eqnarray}
\label{eq:moments-I}
I & = & - \frac{K_I \, e_l}{\mathcal{P}_1(\cos \phi)\mathcal{P}_2(\cos \phi)} 
      \times \\
\nonumber
 & &  \left\{ c_0^I + c_1^I \cos(\phi) + c_2^I \cos(2\phi) + c_3^I \cos(3\phi)
  +  P_l \, \left[ s_1^I \sin(\phi) + s_2^I \sin(2\phi) \right] \right\},
\end{eqnarray}
where $K_I$ is a kinematic factor and $e_l=\pm1$ is the charge of the lepton 
beam with longitudinal polarisation $P_l$. The virtual-lepton propagators 
${\mathcal{P}_{1,2}(\phi)}$ of the BH process introduce an extra 
$\cos \phi$-dependence. The Fourier coefficient 
$c_1^I (s_1^I)$ is proportional to the real (imaginary) part of 
a certain linear combination of the four twist-2 CFFs  
$\mathcal{H},\mathcal{E},\widetilde{\mathcal{H}},\widetilde{\mathcal{E}}$,
the detailed expression depending on the target polarisation (cf. 
Equations~\ref{eq:Mtilde},\ref{eq:LTSA3},\ref{eq:TTSA_N},\ref{eq:TTSA_S}). 
The coefficient $c_0^I$ is related to approximately the same combination 
of CFFs as $c_1^I$, but it is kinematically suppressed by $1/Q$ (Belitsky 
\etal 2002; Diehl 2003). The coefficient $c_1^I$ is sensitive to the D-term 
that was mentioned in Section~\ref{sec:GPDs} (Polyakov and Weiss 1999). The 
coefficients $c_2^I$ and $s_2^I$ describe twist-3 amplitudes and scale as 
$1/Q$, whereas $c_3^I$ is $\alpha_S$-suppressed at leading twist (Diehl 1997). 
In case of a polarised target, additional sums with analogous structure
appear in Equation~\ref{eq:moments-I} (Diehl and Sapeta 2005). In particular,
terms $s_3^I \sin(3\phi)$ appear where $s_3^I$ is sensitive to contributions
from gluon transversity, as in the case of $c_3^I$.
%
%
%
%
\section{Azimuthal Cross Section Asymmetries}
\label{sec:Asyms}
%
%
\subsection{Unpolarised Target}
\label{subsec:UnpolTarget}
%
%
The {\em beam-spin asymmetry} (BSA) for a longitudinally ($L$) 
polarised beam and an {\em unpolarised} ($U$) proton target is defined as
\begin{equation}
\label{eq:BSA1}
A_{LU}(\phi) = \frac
  {d \sigma^{\rightarrow}(\phi) 
 - d \sigma^{\leftarrow}(\phi)}
  {d \sigma^{\rightarrow}(\phi) 
 + d \sigma^{\leftarrow}(\phi)},
\end{equation}
where $\rightarrow$ ($\leftarrow$) denotes beam spin parallel (antiparallel)
to the beam direction. Similarly, the {\em beam-charge asymmetry} (BCA) for an
{\em unpolarised} beam of charge $C$ scattering from an unpolarised proton 
target is defined as:
\begin{equation}
\label{eq:BCA1}
A_{C}(\phi) = \frac
  {d \sigma^+(\phi) - d \sigma^-(\phi)}
  {d \sigma^+(\phi) + d \sigma^-(\phi)},
\end{equation}
where the superscripts $+$ and $-$ denote the lepton beam charge.

Evaluating these asymmetries using Equations~\ref{eq:DVCS-BH-amplitude} and 
\ref{eq:moments-I} to leading power in $1/Q$ in each contribution, and to 
leading order in $\alpha_S$, only the $\sin \phi$  ($\cos \phi$) term 
remains in the numerator of the beam-spin (beam-charge) asymmetry. To the 
extent that the leading $BH$-term $c_0^{BH}$ dominates the denominator, the 
products of the virtual-lepton propagators, 
$\mathcal{P}_1(\cos \phi)\mathcal{P}_2(\cos \phi)$, cancel. In this 
approximation, the azimuthal dependence of the beam-spin
(beam-charge) asymmetry is reduced to $\sin \phi$ ($\cos \phi$):
\begin{equation}
\label{eq:BSA2}
A_{LU}(\phi) \propto \frac{1}{c_{0,U}^{BH}} \, s_{1,U}^I \, \sin \phi
       \propto Im \; \widetilde{M} \, \sin \phi,
\end{equation}
\begin{equation}
\label{eq:BCA2}
A_{C}(\phi) \propto \frac{1}{c_{0,U}^{BH}} \, c_{1,U}^I \, \cos \phi
       \propto Re \; \widetilde{M} \, \cos \phi,
\end{equation}
where the additional subscript (U) of the Fourier coefficients
denotes the unpolarised target. It
appears that both beam-charge and beam-spin asymmetries are sensitive 
to the {\em same} linear combination $\widetilde{M}$ of CFFs which
describes an {\em unpolarised} proton target (Belitsky \etal 2002):
\begin{equation} 
\label{eq:Mtilde}
\widetilde{M} =  \frac{\sqrt{t_0-t}}{2m} \,
                 \left[F_1 \, {\cal H} + \xi \, (F_1 + F_2) \, 
                 \widetilde {\cal H} - \frac{t}{4 m^2} F_2 \, {\cal E}
                \right].
\end{equation}
Here $-t_0=4\xi^2m^2/(1-\xi^2)$ is the minimum  possible value
of $-t$ at a given $\xi$.

Note that, almost independently on details of GPD models, the 
GPDs $H^q$ are expected to dominate this expression, because
i) the second term is suppressed by at least a factor of 10, as 
$\xi$ is usually not larger than 0.2 even in fixed-target kinematics
(cf. Figure~\ref{fig:KinemCoverage}) and the unpolarised
contribution $\mathcal{H}$ is expected to dominate the polarised
one $\widetilde{\mathcal{H}}$, in analogy to the forward case;
ii) the third term is $t$-suppressed, by about a factor of 25 for 
typical $t$-values of about 0.15~GeV$^2$.
For scattering on the proton, the GPD $H^u$ will yield the major
contribution to $\widetilde{M}$ because of $u$-quark dominance.
%
%
\subsection{Polarised Target}
\label{subsec:PolTarget}
%
%
%
In case of a {\em polarised} proton target further sums appear in
Equation~\ref{eq:moments-I}, as mentioned above. They contain other
linear combinations than $\widetilde{M}$, so that measurements of 
{\em target-spin asymmetries} deliver valuable additional experimental 
information.

The single-spin asymmetry w.r.t. to the polarisation of a 
{\em longitudinally} (L) polarised target (LTSA) is defined as
\begin{equation}
\label{eq:LTSA1}
A_{UL}(\phi) = \frac
  {d \sigma^{\Leftarrow}(\phi) 
 - d \sigma^{\Rightarrow}(\phi)}
  {d \sigma^{\Leftarrow}(\phi) 
 + d \sigma^{\Rightarrow}(\phi)}
\end{equation}
where $\Leftarrow$ ($\Rightarrow$) denotes target spin antiparallel (parallel)
to the beam direction. In the above introduced approximation and for
`balanced' beam polarisation ($\left< P_l \right> \approx 0$), its azimuthal
dependence is also purely sinusoidal:
\begin{equation}
\label{eq:LTSA2}
A_{UL}(\phi) \propto \frac{1}{c_{0,L}^{BH}} s_{1,L}^I \, 
             \sin(\phi).
\end{equation}
Neglecting terms of ${\cal{O}}(\xi^2)$ and higher, the Fourier coefficient 
$s_{1,L}^I$ is sensitive to a linear combination of CFFs different from
Equation~\ref{eq:Mtilde} (Belitsky \etal 2002; Diehl 2003):
\begin{eqnarray}
\label{eq:LTSA3}
s_{1,L}^I & \propto & \frac{\sqrt{t_0-t}}{2m} \;
                    \Im \left[ F_1 \widetilde{\mathcal{H}}
                   + \xi\left( F_1+F_2 \right) 
                     \left( \mathcal{H} + \frac{\xi}{1+\xi} \mathcal{E} \right)
                     \right. \nonumber \\
          &         &
           - \left. \left( \frac{\xi}{1+\xi} F_1 + \frac{t}{4m^2} F_2 \right) 
                    \, (\xi \widetilde{\mathcal{E}}) \right].
\end{eqnarray}
$A_{UL}$ is expected to be most sensitive to a combination of $H^q$ and 
$\widetilde{H^q}$, because the kinematic suppression of the second term 
in Equation~\ref{eq:LTSA3}, as compared to the first one, may approximately 
compensate the expected dominance of the unpolarised GPDs $H^q$ over their
polarised counterparts $\widetilde{H^q}$. Hence both should become separable
by combining this measurement with asymmetries measured on an 
unpolarised target. For not too small values of $t$ there exists also 
some sensitivity to $(\xi \, \widetilde{\mathcal{E}})$, which is written
in this way as $\widetilde{\mathcal{E}}$ itself is inversely proportional
to $\xi$ (Goeke \etal 2001; Diehl 2003).

For a {\em transversely} (T) polarised target, the definition of the
single-spin asymmetry (TTSA) $A_{UT}$ is more complicated. The
additional dependence on the azimuthal angle $\phi_S$ of the spin 
vector creates `normal' (N) and 'sideways' (S) 
components. In the approximation used for Equation~\ref{eq:LTSA3}, 
the corresponding Fourier coefficients contain yet further 
combinations of CFFs (Diehl and Sapeta 2005). The normal component reads:
\begin{eqnarray}
\label{eq:TTSA_N}
s_{1,N}^I & \propto & - \frac{t}{4m^2} \; \Im \left[ 
                        F_2 \mathcal{H} - F_1 \mathcal{E} 
            + \xi (F_1+F_2) \, (\xi \, \widetilde{\mathcal{E}}) \right].
\end{eqnarray}
This is known to be the {\em only} combination of CFFs where the GPDs
$E^q$ are not kinematically suppressed as compared to $H^q$. Hence DVCS
measurements on a transversely polarised proton target, in particular
of the normal contribution,  appear to be indispensable for the 
evaluation of the total quark angular momentum through the Ji relation 
(\ref{eq:JiSumRule}). An inherent complication for the measurement
of this relation lies in the fact that both the GPDs $H^q$ and $E^q$ 
need to be measured towards lowest possible values of $t$, while the
$\phi$-dependence of the cross section disappears in the limit 
$t \rightarrow 0$; the relevant asymmetry is suppressed by a factor 
of $\sqrt{-t}/2m$ when extracting the GPDs $H^q$ and even by a factor of 
$t/4m^2$ when extracting the GPDs $E^q$, as can be seen from a comparison of
Equations \ref{eq:Mtilde} and \ref{eq:TTSA_N}.

The sideways component, written in the above used approximation, 
undergoes the same kinematic suppression as the normal component:
\begin{eqnarray}
\label{eq:TTSA_S}
s_{1,S}^I & \propto & - \frac{t}{4m^2} \; \Im \left[ 
           F_2 \mathcal{\widetilde{H}} 
          +  \xi (F_1+F_2) \mathcal{E} 
          - (F_1+\xi F_2) \, (\xi \, \widetilde{\mathcal{E}}) \right].
\end{eqnarray}
It offers access to the imaginary part of a combination of {\em both 
polarised} CFFs, $\widetilde{\cal{H}}$ and $(\xi \, \widetilde{\cal{E}})$, 
although accompanied by the unpolarised CFF $\cal{E}$ whose
$\xi$-suppression will presumably be compensated by its larger size. 
%
%
\subsection{Beyond DVCS}
\label{subsec:BeyondDVCS}
%
%
%
\noindent i) Similar as for differently polarised targets in DVCS, for 
hard exclusive processes other than DVCS, as {\em eg} meson production, 
the involved amplitudes embody other subsets of proton GPDs that are 
linearly combined by different kinematic suppression factors. In 
other words, a certain process is more sensitive to
an individual GPD than another and hence a complete as possible 
determination of GPDs requires measurements of several
hard exclusive reactions and/or final states.

\noindent ii) Every experiment covers a peculiar subspace of the 
$(x_B, t, Q^2)$ phase space, with certain overlap regions between experiments
as will be detailed at the beginning of section~\ref{sec:FutDVCSmeas}.
Hence a certain GPD combination accessible
through a certain cross section, cross section difference or cross 
section asymmetry will be surveyed by different 
experiments in partly overlapping subspaces only.

\noindent iii) In `associated' DVCS, where the proton target does not 
stay intact, the formalism of angular analysis remains the same, while 
the accessible GPDs  are different from those of the proton (Diehl 2003).

\noindent iv) Hard exclusive leptoproduction on nuclear targets proceeds 
either coherently, {\em ie}, by scattering on the nucleus as a whole, 
or incoherently, {\em ie} on a single proton or neutron. Coherent scattering
proceeds preferentially at (very) small values of $t$. For small and 
medium values of $t$, the electromagnetic form factor of the neutron is 
small, leading to a small Bethe-Heitler cross section, as compared to DVCS. 
Hence in this $t$-region the interference term $I$ is suppressed  for 
scattering on the neutron and incoherent nuclear DVCS is expected to 
behave similarly to DVCS on the proton.

\noindent v) Gluon GPDs can be accessed in DVCS and vector meson production 
(Diehl 2003; Goloskokov and Kroll 2005), in particular at small $\xi$, 
{\em ie} at collider kinematics, but also in $\phi$-production at 
fixed-target kinematics (Diehl and Vinnikov 2004).

%
%
%
\section{Experimental Results on DVCS}
\label{sec:ExpResults}
%
\subsection{Collider Experiments}
\label{subsec:ColliderResults}
%
The DVCS cross section has been measured in hard exclusive photon
electroproduction at the {\sc Hera} collider by the experiments 
{\sc H1} and {\sc Zeus}. It becomes accessible by integrating 
Equation~\ref{eq:DVCS-BH-amplitude} over its azimuthal dependence (see
section~\ref{subsec:CFFs}).
For the $x_B$-range accessible at collider kinematics
(10$^{-2}$ ... 10$^{-4}$), two-gluon exchange plays a major role 
besides the above discussed quark-exchange handbag diagram, 
{\em ie, both} gluon and quark GPDs are probed simultaneously
but only for very small skewedness values below 10$^{-2}$. 

The analysis method used is similar in both experiments due to their 
similar geometry. As the outgoing proton
remains undetected in the beam pipe, the event topology is
defined by two electromagnetic clusters, the outgoing lepton and 
the produced real photon, and at most one associated track. 
Two events samples are selected:

\noindent i) in the `DVCS-enriched' sample a hard, {\em ie} centrally 
produced real photon is required, while the outgoing lepton is 
measured under a small scattering angle w.r.t. the incoming one; still 
a high enough virtuality $Q^2 > 4$~GeV$^2$ is ensured by requiring a large
energy of the scattered lepton ($> 15$~GeV).

\noindent ii) in the Bethe-Heitler dominated `reference sample' the 
radiatively produced photon is emitted under a small angle w.r.t. the incoming 
lepton, and the outgoing lepton is measured in the central region. 

A Monte Carlo simulation of the completely known BH process, which describes 
the reference sample, is used to subtract the BH contribution from 
the DVCS-enriched sample. The remainder of the spectrum is due to DVCS and 
possible additional background; no contribution from the interference term 
exists at leading twist, as the data are integrated over the azimuthal
angle. The  $Q^2$-dependence of the differential 
$\gamma^* \, p \rightarrow \, \gamma \, p$ cross section is illustrated in 
%
\begin{figure}[htb]
\label{fig:H1+GPDvsQ2}
\includegraphics[width=10cm]{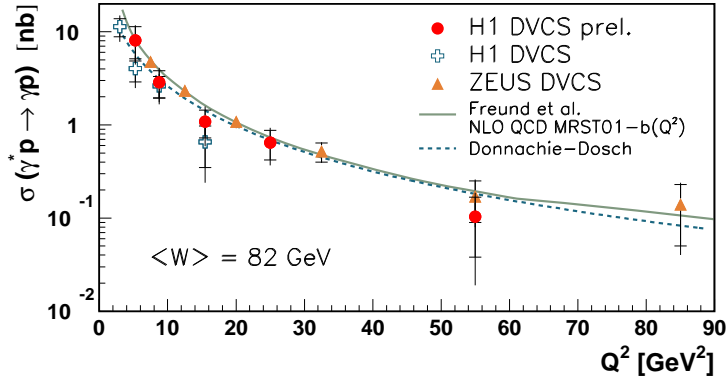}
\centering
\caption{\it $Q^2$-dependence of the differential 
$\gamma^* \, p \rightarrow \gamma \, p$ cross section measured by {\sc H1} 
and {\sc Zeus} in comparison to a GPD-based NLO pQCD calculation. For 
comparison, a prediction of a Colour Dipole model is also shown.
The Figure is taken from (Favart 2004).}
\end{figure}
Figure~4. The data shown are from {\sc H1}, both earlier published 
(Adloff \etal 2001) and recent preliminary (Favart 2004), and from 
{\sc Zeus} (Chekanov \etal 2003), the latter based on substantially higher 
statistics. The solid curve shows a NLO pQCD calculation 
(Freund and McDermott 2002) using a GPD parameterisation based on 
MRST2001 PDFs and a $Q^2$-dependent $t$-slope $b(Q^2)$ describing the
factorised $t$-dependence (Freund \etal 2003). 
Using instead CTEQ6 PDFs with the same $t$-slope (not shown) yields
a very similar $Q^2$-dependence, but a different normalisation. For 
comparison a Colour Dipole model calculation (Donnachie and Dosch 2001) 
is also shown in the Figure. Agreement between all data sets 
and models can be seen in the log-scale representation of the 
$Q^2$-dependence, although there seem to be discrepancies at lower 
values of $Q^2$.
\begin{figure}[htb]
\label{fig:H1+ZEUSvsW}
\includegraphics[width=10cm]{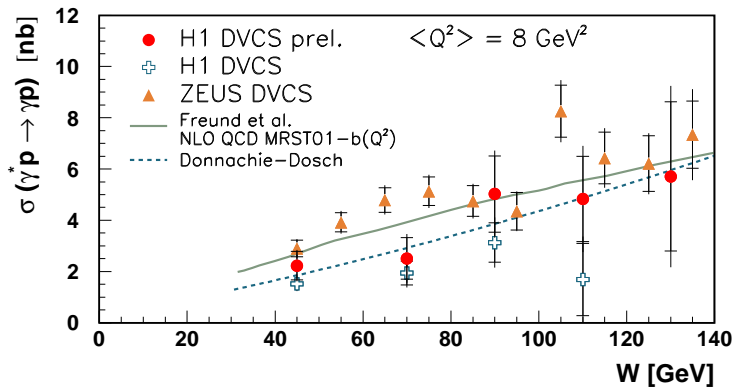}
\centering
\caption{\it $W$-dependence of the differential 
$\gamma^* \, p \rightarrow \gamma \, p$ cross section measured by H1 and ZEUS,
compared to a GPD-based NLO pQCD calculation. For comparison, a prediction of
a Colour Dipole model is also shown . The Figure is taken from 
(Favart 2004).}
\end{figure}

The corresponding $W$-dependence is displayed in Figure~5
for the same data sets and models, where $W$ is the center-of-mass energy of 
the system of virtual photon and proton. The virtuality appears high
enough to assign the observed steep rise with $W$ to the nature of DVCS as 
a hard process, as increasing $W$ implies decreasing $x_B$, where the parton 
densities in the proton show a fast rise. Most data reside at lower $Q^2$
($\langle Q^2 \rangle=8$~GeV$^2$), the region of possible discrepancies 
between data sets. Presently a 2$\sigma$ difference exists between 
{\sc Zeus} and {\sc H1} data in the medium $W$-range. Differences of similar
size exist between different model calculations. Note that, since the slope 
of the $t$-dependence of GPDs is still unknown, the normalisation of the 
GPD-based curves shown above remains arbitrary to some extent. Future 
measurements of the $t$-dependence of the DVCS cross section are therefore 
of high importance.
%
%
%
%
\subsection{Fixed-target Experiments}
\label{subsec:Fixed-targetResults}
%
In sections~\ref{sec:DVCS} and \ref{sec:Asyms} it was shown that in hard 
exclusive real-photon leptoproduction the
interference of the Bethe-Heitler and Deeply Virtual Compton Scattering
processes is a rich source for extracting a wealth of information on GPDs. 
In fact, the first published GPD-related experimental results were 
beam-spin asymmetries measured in DVCS on the proton by the 
fixed-target experiments {\sc Hermes} at {\sc Hera} (Airapetian \etal 2001) 
with a positron beam and by {\sc Clas} at Jefferson Laboratory 
(Stepanyan \etal 2001) with an electron beam. 
\begin{figure}[htb]
\label{fig:CLAS_BSAvsPhi}
\noindent
\begin{minipage}[b]{.49\linewidth}
  \centering \includegraphics[width=\linewidth]{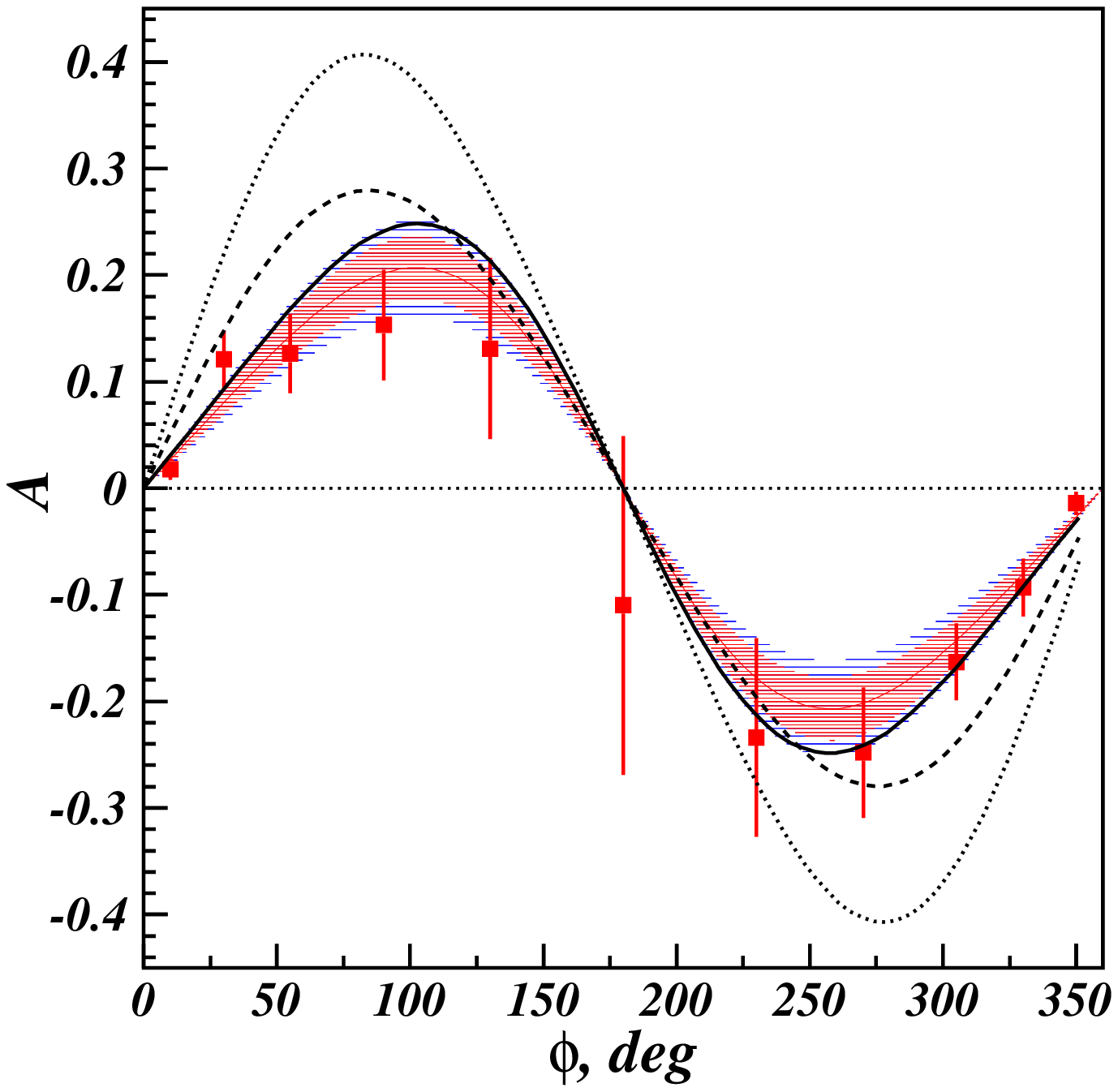}
\end{minipage} \hfill
\begin{minipage}[b]{.49\linewidth}
  \centering \includegraphics[height=6cm]{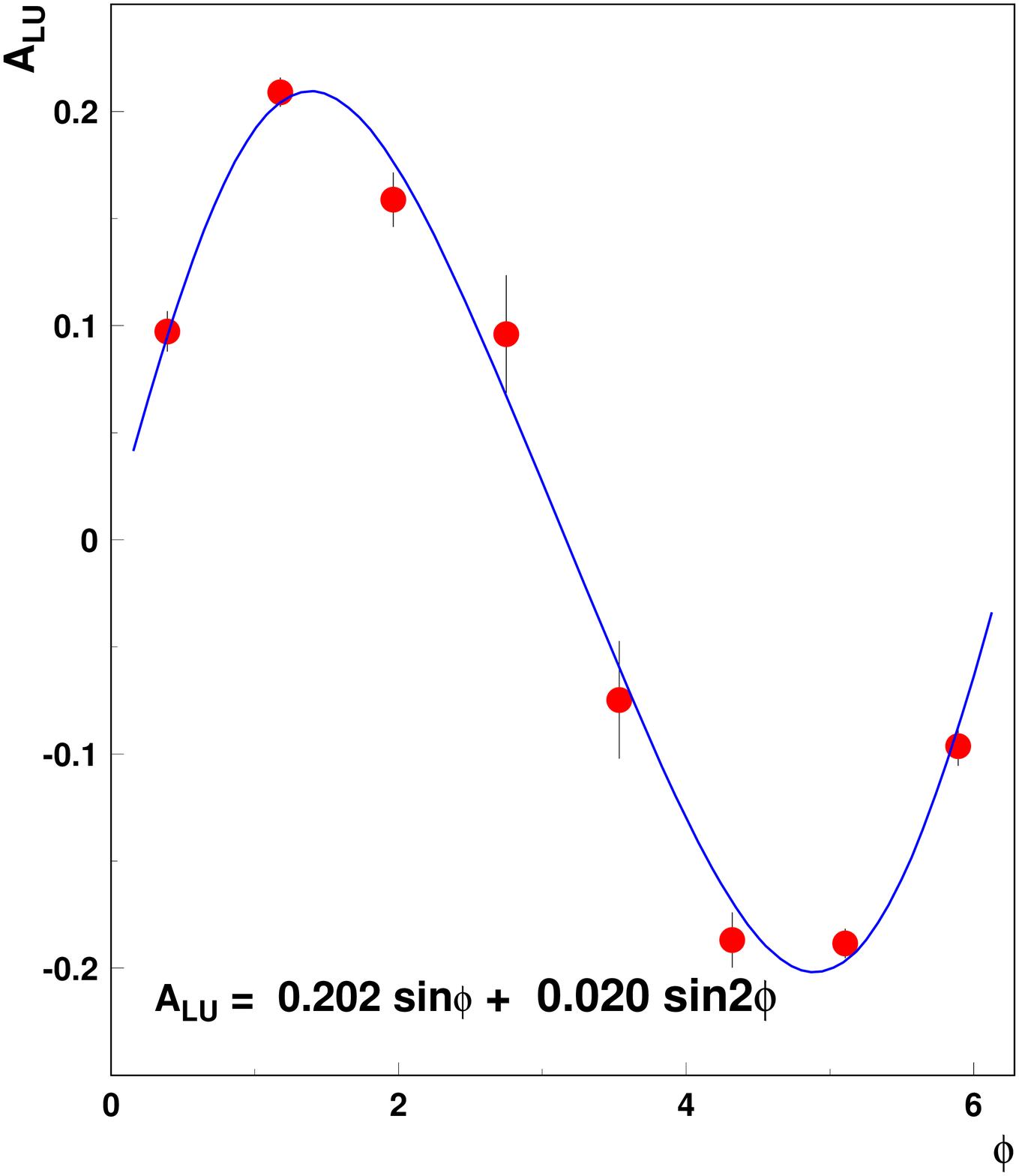}
\end{minipage}
\centering
\caption{\it {\sc Clas}: Azimuthal dependence of the beam-spin asymmetry.
        Left: earlier data at 4.25 GeV. Right: recent preliminary data at 
        5.75 GeV. Only statistical errors are shown.}
\end{figure}
\begin{figure}[htb]
\label{fig:HERMES_BSAvsPhi}
\noindent
\begin{minipage}[b]{.49\linewidth}
  \centering \includegraphics[width=\linewidth]{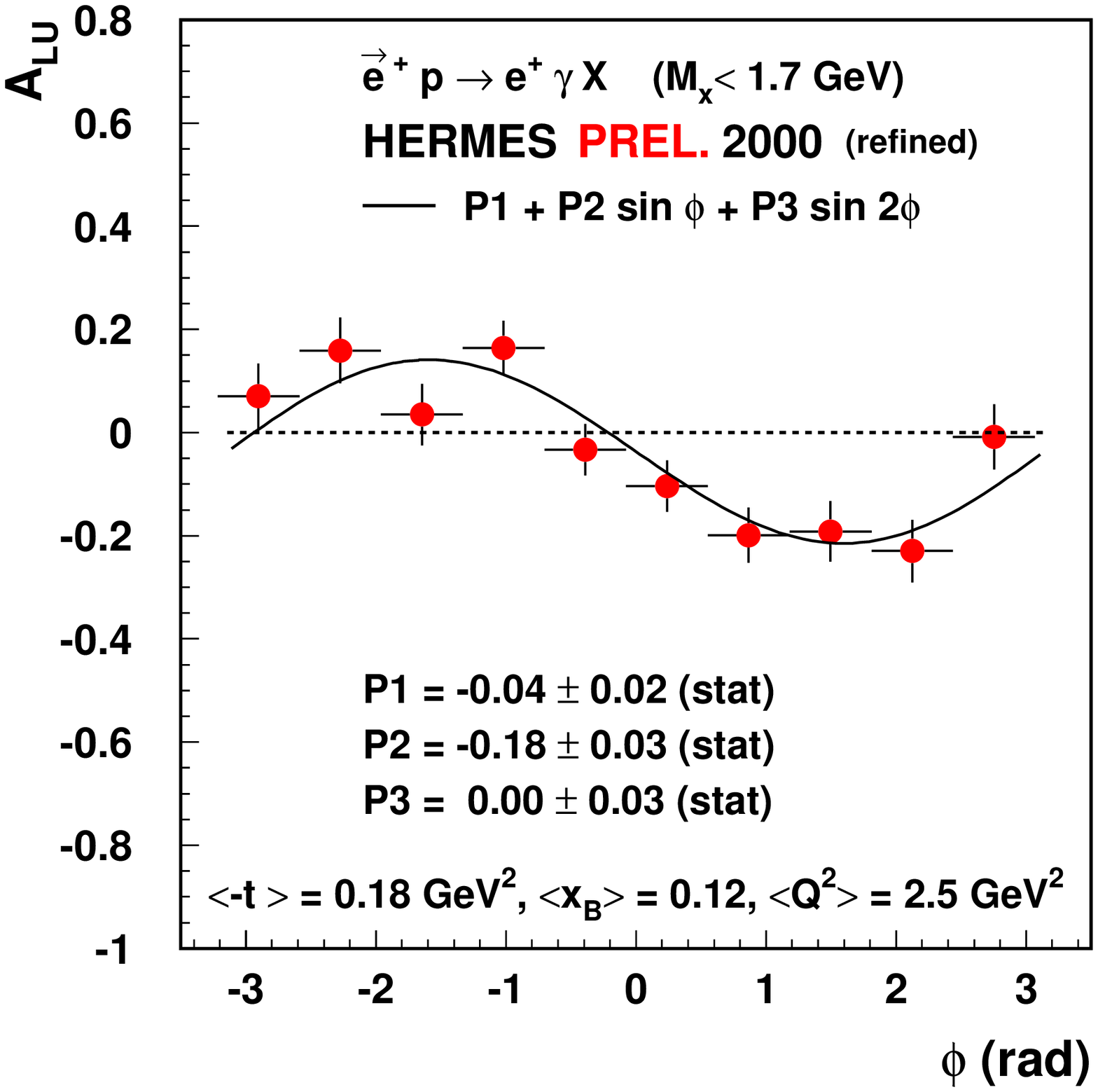}
\end{minipage} \hfill
\begin{minipage}[b]{.49\linewidth}
  \centering \includegraphics[width=\linewidth]{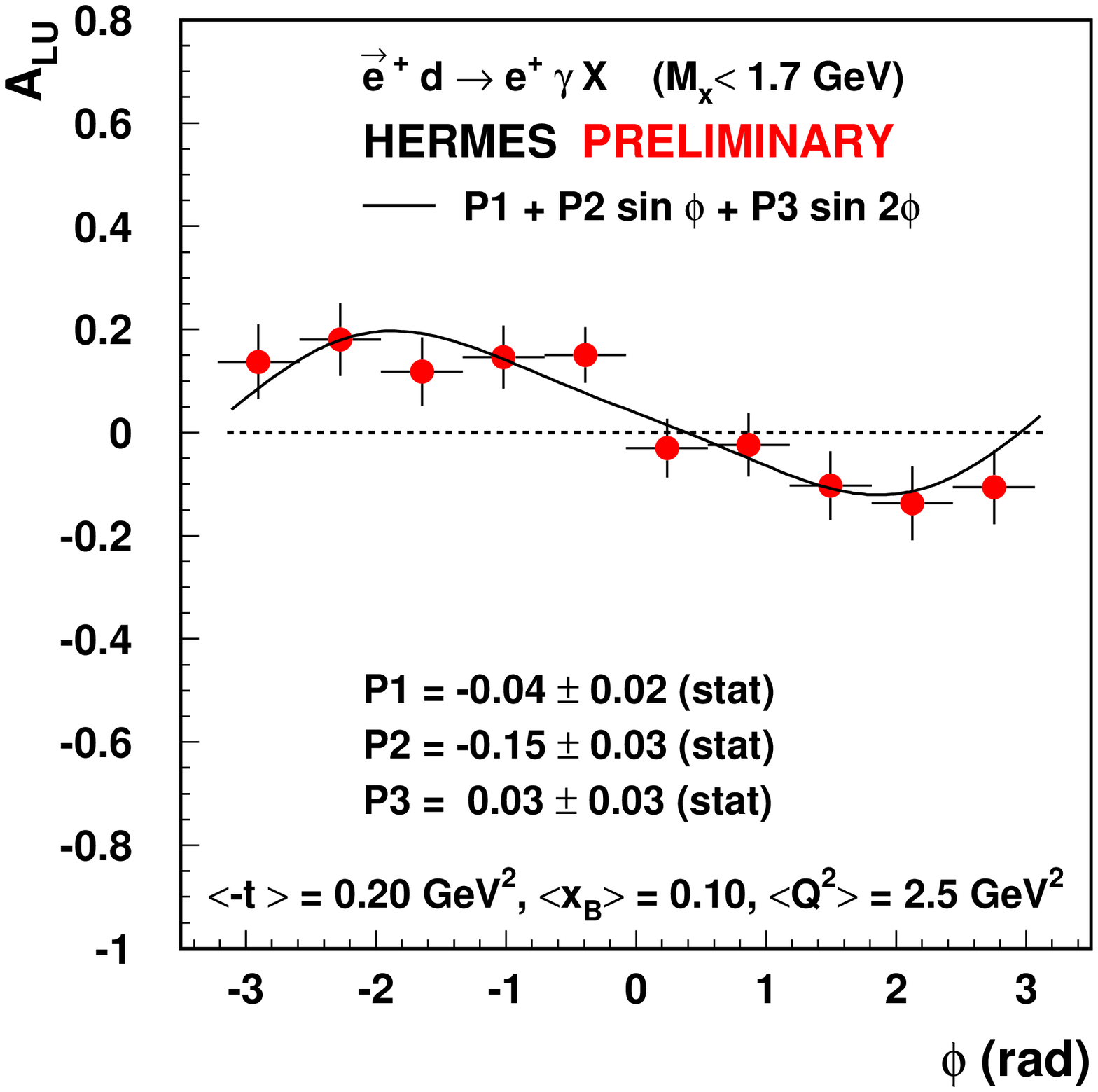}
\end{minipage}
\centering
\caption{\it {\sc Hermes}: Azimuthal dependence of the beam-spin asymmetry
             on proton (left) and deuteron (right), measured at 27.6 GeV. 
             Only statistical errors are shown.}
\end{figure}

Note that opposite beam charges mean opposite signs of the measured 
BSAs, a fact that is not apparent when comparing Figures 6 and 7
due to different $\phi$-ranges chosen.
Meanwhile, more precise (preliminary) BSA measurements were presented 
by both experiments. 

The average beam polarisation at {\sc Clas} ({\sc Hermes}) was 70\% (55\%).
Both experiments use the missing-mass technique to compensate for present
incompletenesses of their detectors; at {\sc Clas} ({\sc Hermes}) the
real photon (recoiling proton) remains undetected. At {\sc Clas}, 
the missing-mass resolution cannot cleanly separate the 
$ep\pi^0$ from $ep\gamma$ reactions. The electromagnetic calorimeter detects
only photons above 8$^o$, {\em ie}, it misses most of the DVCS photons 
but detects usually one of the two $\pi^0$ decay photons. In the analysis 
of the 5.75 GeV data, background from $\pi^0$  decay is reduced by applying a 
corresponding veto. A cut $\theta_{\gamma \gamma*}<120$ mrad is used 
to select data at low $t$. The missing mass squared 
for the undetected real photon is restricted to $M_X^2<0.025$ GeV$^2$. 
The kinematics coverage for $W>2$ GeV is $1.2<Q^2<4$ GeV$^2$ and
$0.1<x_B<0.5$, and the analysis is restricted to $-t<0.5$ GeV$^2$.
At {\sc Hermes}, to account for the limited resolution of the spectrometer,
an asymmetric missing-mass interval around the proton mass is chosen 
(called `exclusive bin': $-1.5<M_X<1.7$ GeV), based on signal-to-background 
studies using a Monte Carlo simulation. Kinematic requirements to the 
outgoing lepton are $1<Q^2<10$ GeV$^2$, $W^2>9$ GeV$^2$ and $\nu<22$ 
GeV, implying $0.03<x_B<0.35$. For the results shown in Figure~7
(Ellinghaus \etal 2002b), the polar angle between virtual and produced real 
photon obeys $2<\theta_{\gamma \gamma*}<70$ mrad. Monte Carlo studies show 
that the exclusive sample contains about 10\% associated events, where the 
nucleon doesn't stay intact, and about 5\% events from DIS fragmentation. 
Note that for the exclusive sample the variable $t$ is calculated assuming the 
3-particle final state $ep\gamma$, thereby considerably improving the 
$t$-resolution, and the analysis is restricted to $-t<0.7$ GeV$^2$.

{\sc Clas} proton data for average kinematics 
$\left( \right. \left<Q^2 \right>=1.25$ GeV$^2$,$\; \left<x_B \right>=0.19$,
$\left< -t \right>=0.19$ GeV$\left.^2 \right)$ are shown in 
Figure~6, the earlier (Stepanyan \etal 2001) BSA result in the left panel
and the more recent preliminary result (Smith 2003) in the right one. 
The most recent (preliminary) BSA results from {\sc Hermes}, 
on both proton and deuteron (Ellinghaus \etal 2002b), are shown in Figure~7.
The average kinematics  are $\left<Q^2 \right>=2.5$ GeV$^2$,
$\left<x_B \right> \simeq 0.10$ and $\left< -t \right> \simeq 0.20$ GeV$^2$.
All BSA data exhibit 
substantial sinusoidal asymmetries in accordance with the expectation 
given in Equation~\ref{eq:BSA2}. The magnitude of the $\sin \phi$ component was
fitted as $0.202\pm0.028_{stat}\pm0.013_{sys}$ and 
$-0.23\pm0.04_{stat}\pm0.03_{sys}$ for the published data from {\sc Clas} 
and {\sc Hermes}, respectively, while $0.202$ and $-0.18\pm0.03_{stat}$
were obtained from their recent preliminary data. The next-higher
harmonic ($\sin (2\phi)$, see Equation~\ref{eq:moments-I}) is 
found to be compatible with zero within the total experimental uncertainty 
in both experiments. No difference is seen when comparing the {\sc Hermes}
BSA results for proton and deuteron, which is not surprising when recalling 
the argument made in Note iv) of section~\ref{subsec:BeyondDVCS}.
No published data exist yet for kinematic dependences of BSAs. Unpublished 
{\sc Hermes} data (Ellinghaus 2004a) do not show any clear dependence on
$t$, $x_B$ or $Q^2$ within experimental uncertainties. 

A beam-charge asymmetry measurement requires data for both beam charges.
{\sc Hera} is presently the only GeV-range accelerator that provides both 
electron and posi\-tron beams. It offers the additional flexibility of 
switching from time to time, for the same charge of the beam, the direction 
of its polarisation to reduce systematic effects. 
\begin{figure}[htb]
\label{fig:BCAvsPhi}
\begin{minipage}[b]{.49\linewidth}
  \centering \includegraphics[width=\linewidth]{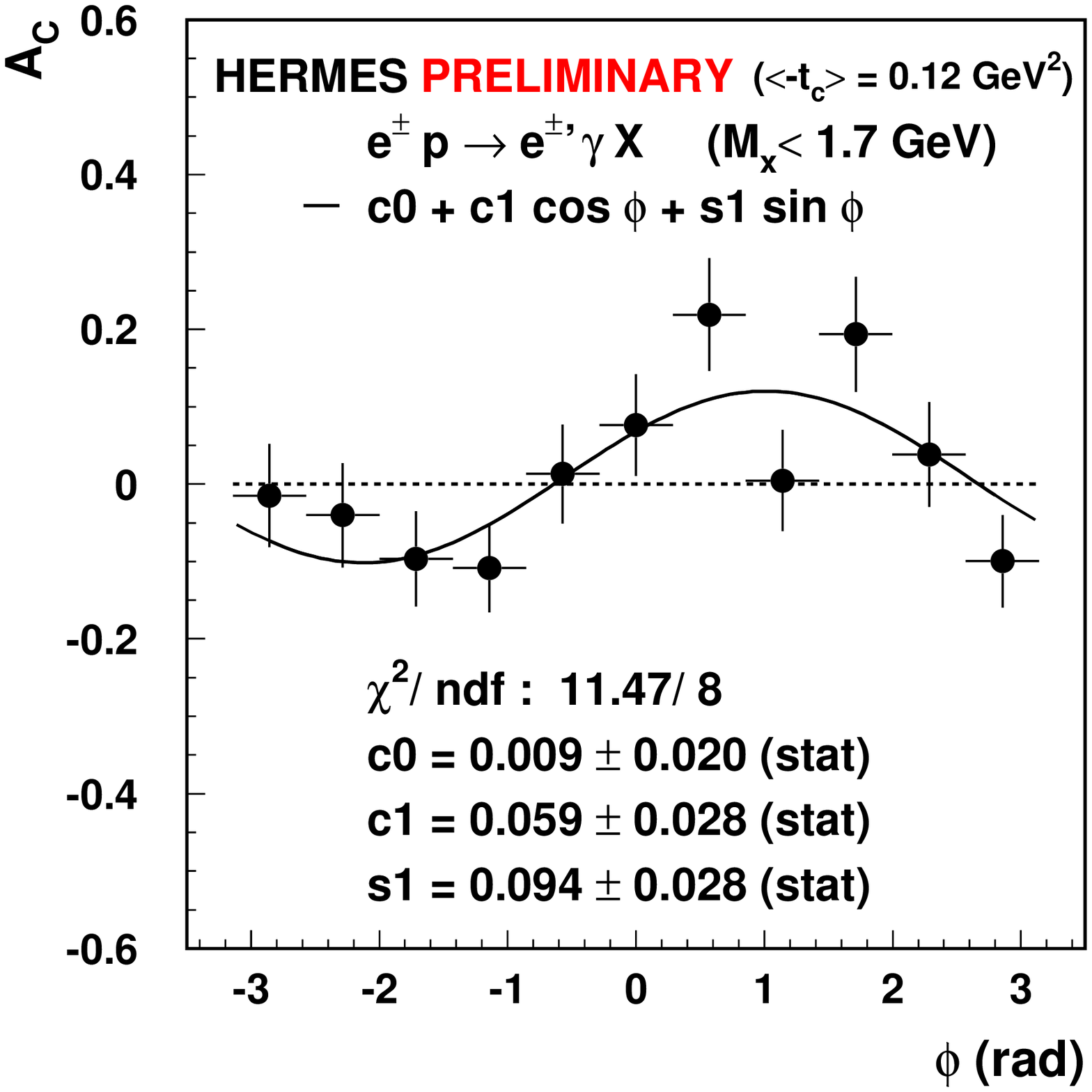}
\end{minipage} \hfill
\begin{minipage}[b]{.49\linewidth}
  \centering \includegraphics[width=\linewidth]{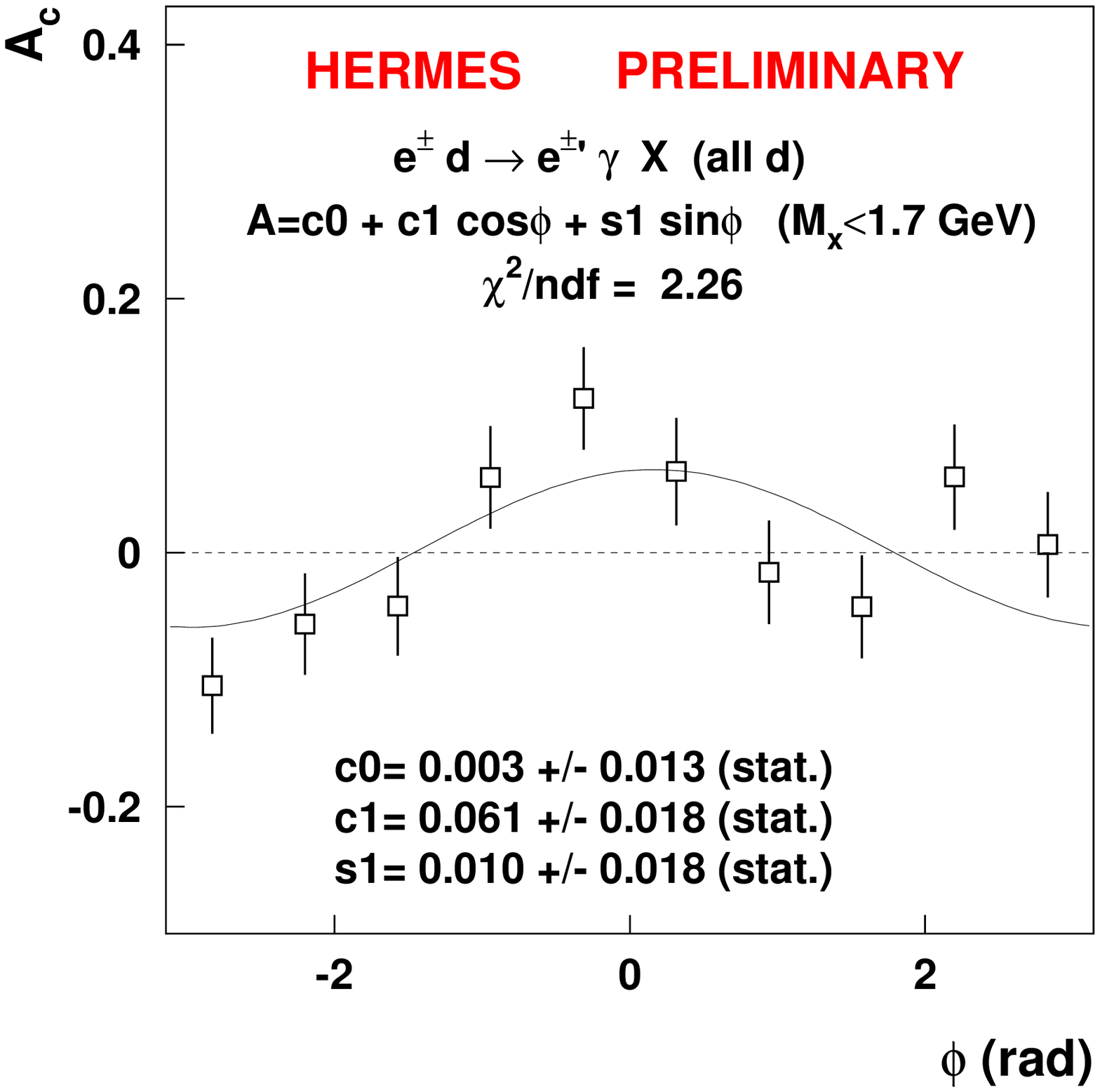}
\end{minipage}
\centering
\caption{\it HERMES: Azimuthal dependence of the beam-charge asymmetry
on proton (left) and on unpolarised and vector-polarisation-balanced 
deuteron (right). Only statistical errors are shown.}
\vspace*{3mm}
\end{figure}
\begin{figure}[htb]
\vspace*{4mm}
\label{fig:BCAvs_t}
\includegraphics[width=7cm]{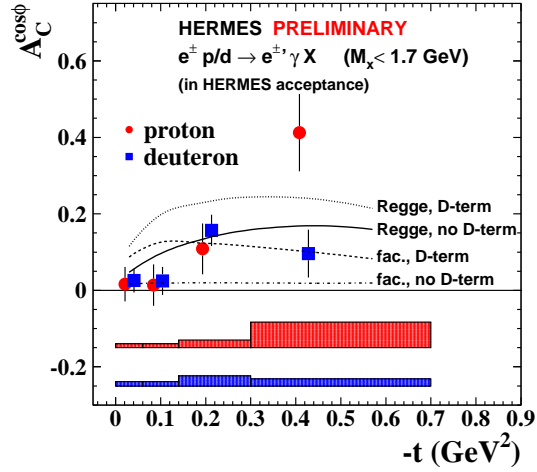}
\centering
\caption{\it HERMES: $t$-dependence of the $\cos \phi$ component 
of the beam-charge asymmetry on proton and deuteron. Statistical (systematic)
uncertainties are indicated by error bars (bands). The curves represent LO 
pQCD calculations using different GPD ans\"atze, see text.}
\vspace*{-3mm}
\end{figure}

In Figure~8 
preliminary BCA results from {\sc Hermes} are shown (Ellinghaus 2004c), 
which were obtained from the same analysis as described above using
$5 < \theta_{\gamma \gamma^*} < 45$ mrad. Both
proton and deuteron data exhibit the expected $\cos \phi$-dependence 
with a similar sizeable magnitude, $0.059\pm0.028_{stat}$ and 
$0.061\pm0.018_{stat}$, again not a surprising agreement. 
The proton BCA also contains a 
significant $\sin \phi$-component that is caused by average beam 
polarisation values not vanishing individually for each beam charge.
For clarity it has to be noted that an about twice as large deuteron BCA 
(not shown here) was reported earlier (Ellinghaus 2002a), obtained by
discarding the low-$t$ region with the requirement 
$15<\theta_{\gamma \gamma*}<70$ which led to a larger average value of 
$\left< -t \right> \simeq 0.27$ GeV$^2$. Both results are consistent 
because the BCA decreases with decreasing $-t$, as will be shown below.

Kinematic dependences are obtained
by subdividing the data set into several bins,  depending on the 
statistics for a given variable at a time.
In each bin a fit of the azimuthal dependence is performed, including the 
same harmonics as indicated in the panels of Figure~8. 
The $t$-dependence of proton and deuteron BCA is shown in Figure~9,
also obtained for $5 < \theta_{\gamma \gamma^*} < 45$ mrad
(Ellinghaus 2004c). As expected, the signal becomes only sizeable from
medium values of $-t$ on. Here proton and deuteron data agree, as discussed 
in Note iv) of section~\ref{subsec:BeyondDVCS}. Incoherent scattering on 
the neutron may become a substantial contribution at larger $-t$-values,
{\em ie} in the last $t$-bin,
and compensate a further increase of the deuteron asymmetry.
No effects are seen from coherent scattering on the deuteron bound state 
which would be present
in the lowest $-t$-bin only. Superimposed to the experimental data are curves 
representing theoretical calculations (Vanderhaeghen \etal 2001) based on 
different GPD models (Vanderhaeghen \etal 1999). They are calculated at
{\sc Hermes} kinematics, separately for the average kinematics in each 
individual bin (Ellinghaus 2004a). On the basis of the available
statistics, the data seem to favour the model with the Regge ansatz
and no D-term contribution. From Figure~9
it can already be concluded, and it will be discussed in more detail in 
section~\ref{subsec:FutHermes}, that
BCA measurements possess a considerable discriminative power against
different ans\"atze and parameterisations in GPD models.
Note that no dependence on $x_B$ or $Q^2$ is seen in unpublished BCA results 
(Ellinghaus 2004a).

\begin{figure}[htb]
\label{fig:NuclearDVCS}
\begin{minipage}[b]{.49\linewidth}
  \centering \includegraphics[width=\linewidth]{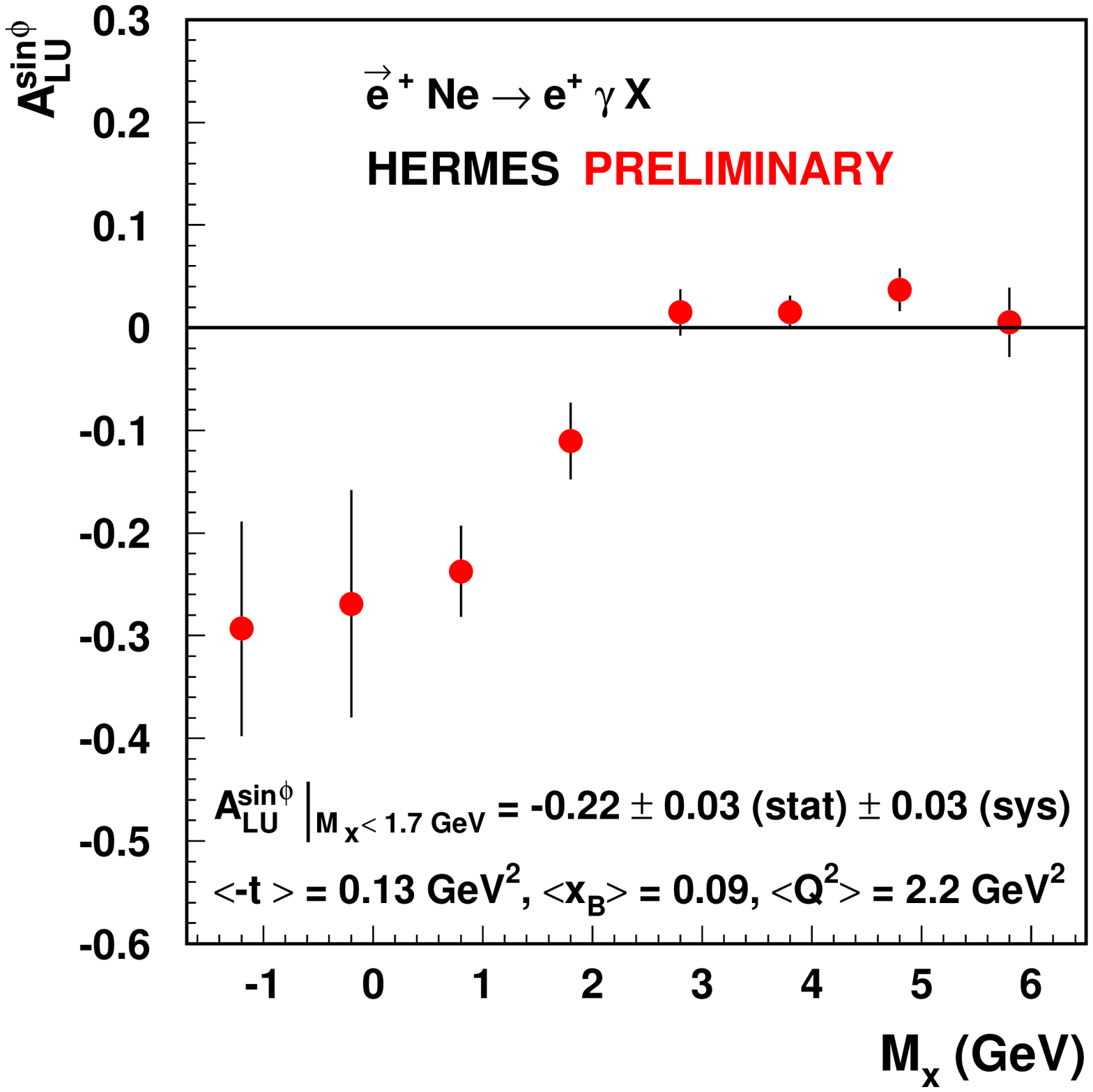}
\end{minipage} \hfill
\begin{minipage}[b]{.49\linewidth}
  \centering \includegraphics[width=\linewidth]{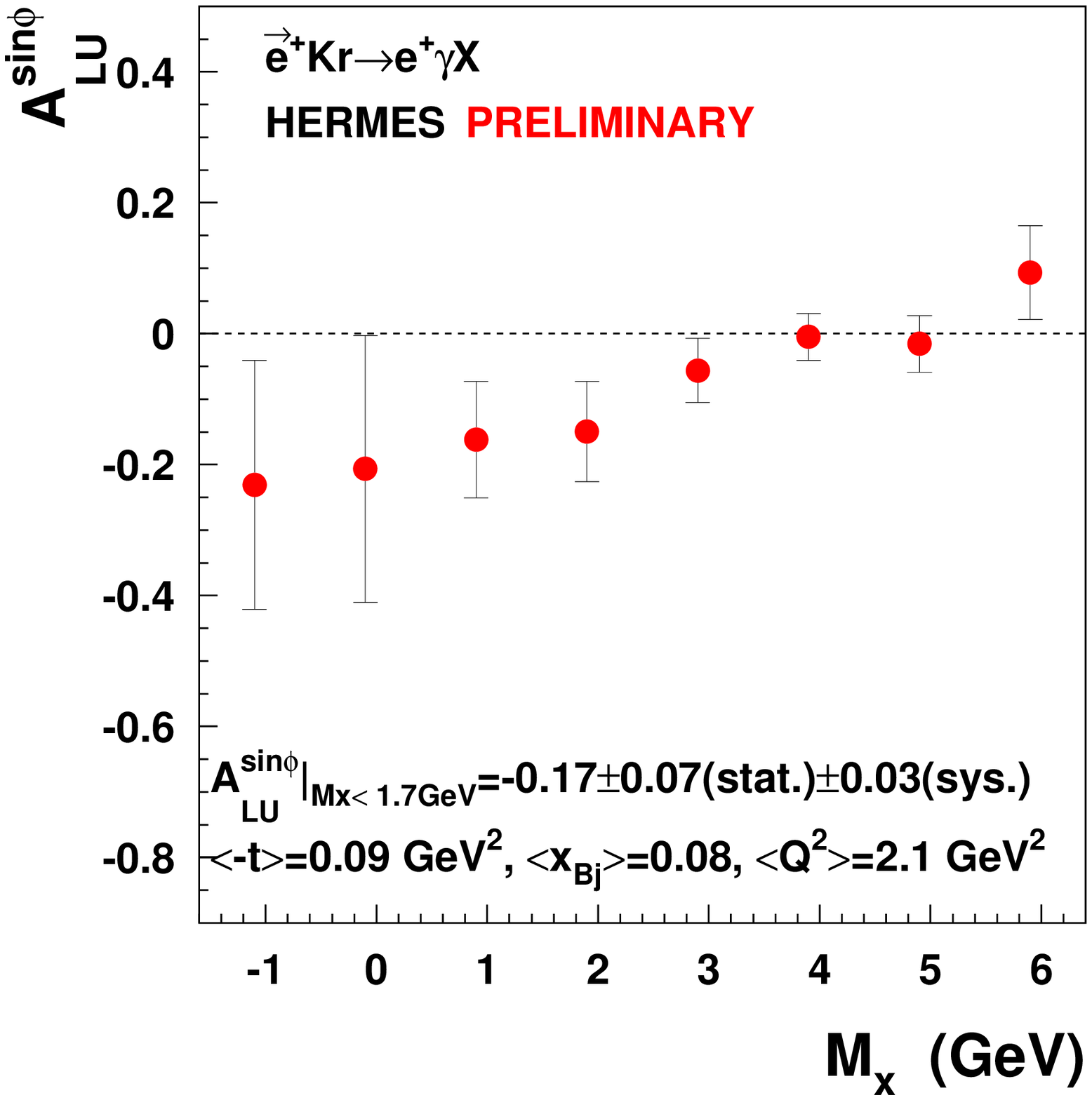}
\end{minipage}
\centering
\caption{\it HERMES: The $\sin (\phi)$-component of the beam-spin 
asymmetry on Neon and Krypton, shown in dependence on missing mass. 
Only statistical errors are given.}
\end{figure}
By measuring DVCS on a longitudinally polarised {\em deuteron} target, 
{\sc Hermes} obtained a preliminary result on the {\em longitudinal 
target-spin} asymmetry (LTSA, cf. Equations~\ref{eq:LTSA1}-\ref{eq:LTSA3}).
Fits as explained above, including $\sin \phi$ and $\sin(2\phi)$ harmonics,
yield asymmetries compatible with zero; both components are found to be
smaller than 0.03 with the total experimental uncertainty being of the 
same order.

DVCS on nuclear targets was briefly mentioned in Note iv) 
of section~\ref{subsec:BeyondDVCS}. Both experimental and theoretical 
information is 
scarce, especially for targets with an atomic number higher than that of
deuterium. In Figure~10,
preliminary BSA results of {\sc Hermes} using Neon and Krypton targets
are shown in dependence on the missing mass, using the proton mass to
calculate the kinematic variables. The average kinematics are indicated 
in the panels, based on $2 < \theta_{\gamma \gamma^*} < 70$ mrad.
As for all {\sc Hermes} results discussed above, sizeable 
asymmetries appear only in the exclusive bin around the target
(proton) mass, while they generally vanish at higher masses. 
It is clearly seen for both Neon and Krypton that already without 
separation of coherent and incoherent processes significant BSAs exist 
in the exclusive bin, while their interpretation can be attempted only 
after the separation. The fitted
size of the $\sin \phi$-component is $-0.22\pm0.03$ ($-0.17\pm0.07$) for
Neon (Krypton), without significant higher harmonics (Ellinghaus \etal 2002b).
For the case of coherent hard exclusive processes on nuclei it was pointed
out that information about the energy, pressure, and shear forces
distributions inside nuclei will become accessible (Polyakov 2003).
%
%
%
\section{Future DVCS Measurements}
\label{sec:FutDVCSmeas}
%
%
%
In Figure~11 
kinematics coverages are compared for DVCS measurements by 
existing or planned fixed-target experiments at {\sc Cern}, {\sc Hera} 
and {\sc JLab}. The kinematic limits are taken from 
(d'Hose 2002), (Ellinghaus 2004a and 2004b) 
and (Cardman \etal 2001). 

\begin{figure}[htb]
\vspace*{-5mm}
\includegraphics[width=8.5cm]{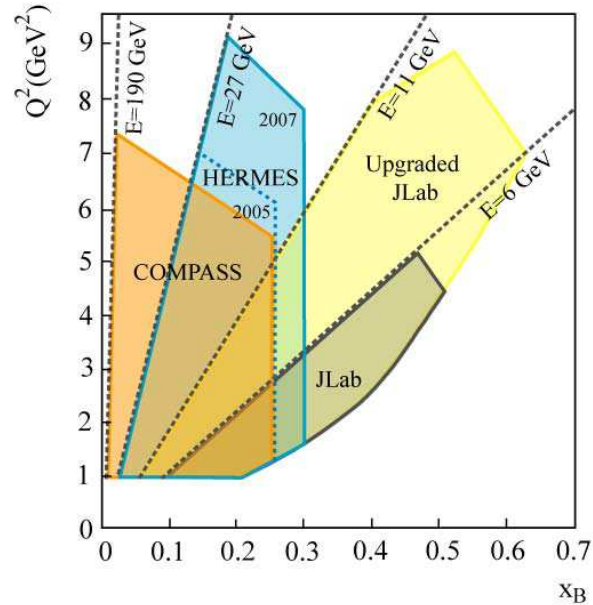}
\centering
\vspace*{-5mm}
\caption{\it Kinematics coverage for fixed-target experiments:
i) {\sc Compass} at 190 GeV; ii) {\sc Hermes} at 27.6 GeV, dotted line for 
existing data ($\leq$ 2005), solid line for future (2005-2007) data 
with an integrated luminosity higher by about one order of magnitude;
iii) {\sc JLab} experiments at 6 GeV (now), and at 11 GeV (after upgrade).}
\label{fig:KinemCoverage}
\end{figure}
As can be seen, the $(x_B, Q^2)$-regions of these
fixed-target experiments do partly overlap, while in comparison to the 
collider experiments at {\sc Hera} there is no overlap in $x_B$ 
(fixed-target above 0.03, collider below 0.01) and only very little overlap 
in $Q^2$ (1...8 GeV$^2$ vs. 5...100 GeV$^2$). Higher $x_B$-values 
($>0.3$) can only be accessed at {\sc JLab}, an advantage of their
relatively low beam energy. At moderate $x_B$, higher $Q^2$-values 
($\simeq$ 8 GeV$^2$) are reachable in the short-term at {\sc Hermes} only. 
Later on, 
the upgraded {\sc JLab} will be able to also reach this region, by 
compensating their lower beam energy by a huge luminosity planned to 
be several orders of magnitude higher than that at other facilities.
%
%
\subsection{{\sc Hera} Collider Experiments}
\label{subsec:HeraColl}
%
%
No published projections exist, to what extent the recent detector upgrades of
the {\sc Hera} collider experiments {\sc H1} and {\sc Zeus} will be beneficial 
to the
ongoing and future measurements of DVCS, until the foreseen shutdown of the
{\sc Hera} accelerator in the middle of 2007. The newly installed spin
rotators make the polarised beam also available to {\sc H1} and {\sc Zeus}.
In both experiments microvertex detectors have been installed which will
allow the precise measurement of the outgoing lepton track, and hence of the 
event vertex, so that the azimuthal angle of the photon can be determined
with higher precision. Altogether, these upgrades will make it possible to
also measure the azimuthal dependence of beam-spin and beam-charge asymmetry
at collider kinematics. It remains to be shown, to what extent these future
data sets will allow the determination of quark or gluon GPDs in the
region of very small $\xi$.
%
%
\subsection{Experiments at Jefferson Lab}
\label{subsec:FutJLab}
%
Jefferson National Laboratory {\sc (JLab)} has approved two dedicated 
DVCS experiments to run at the 6~GeV longitudinally polarised electron 
beam with high luminosity. The first one, the high-resolution arm 
spectrometer E00-110 in Hall~A (Bertin \etal 2000) is  using both 
hydrogen and deuterium targets and finished data taking at the end of 2004.
\begin{figure}[htb]
\label{fig:Fut_CLAS6GeV_BCA_H_vsPhi+t}
\vspace*{5mm}
\begin{minipage}[b]{.49\linewidth}
  \centering \includegraphics[width=\linewidth]
                             {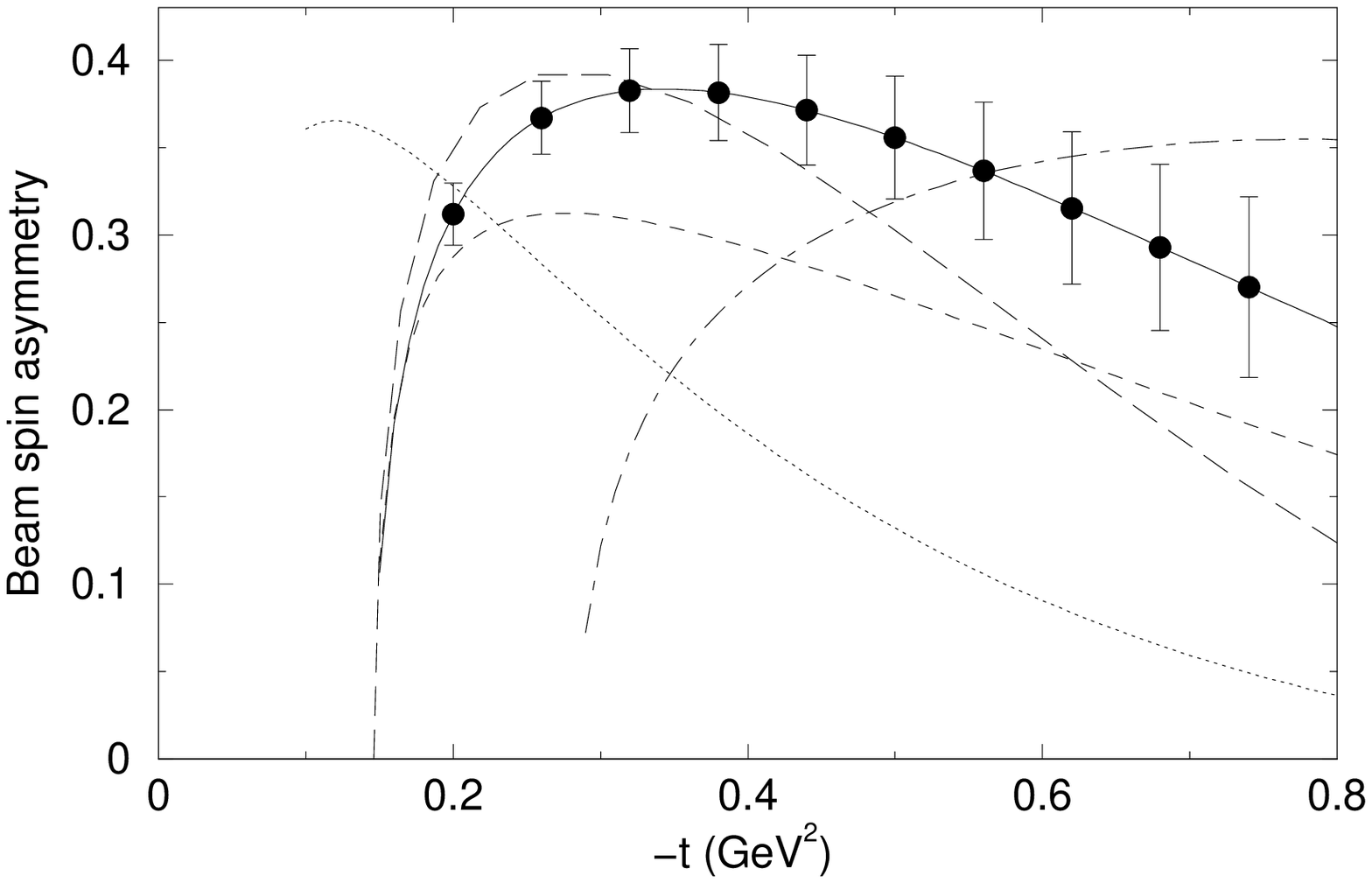}
\end{minipage} \hfill
\begin{minipage}[b]{.49\linewidth}
  \centering \includegraphics[width=\linewidth]
                             {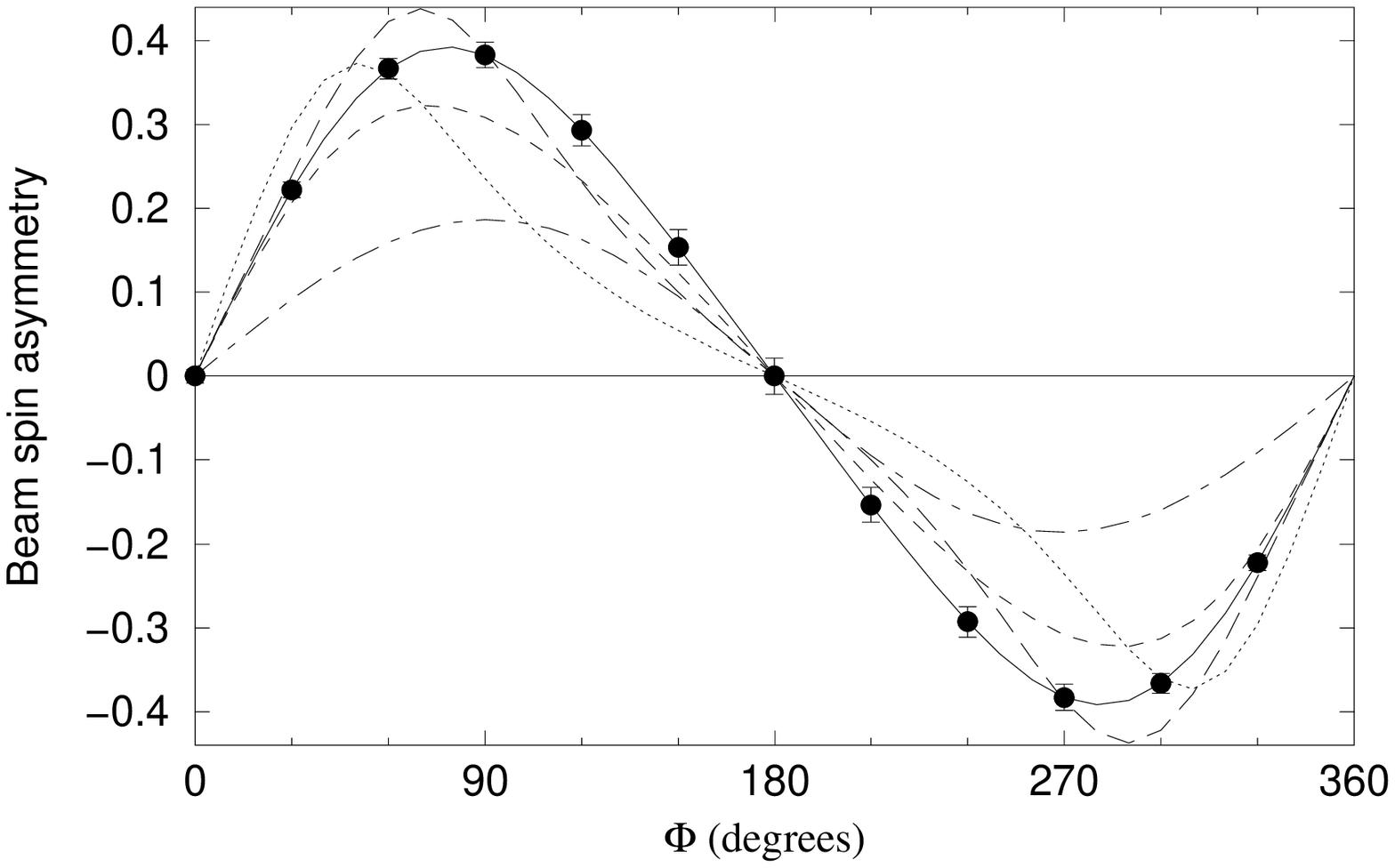}
\end{minipage}
\centering
\caption{\it {\sc Clas}: Projections for beam-spin asymmetries at 6 GeV:
$t$-dependence at $\phi=90^o$ (left) and $\phi$-dependence at 
$-t = 0.325$ GeV$^2$ (right). Projected statistical errors are given
at $Q^2 = 2\pm0.5$ GeV$\, ^2$ and $x_B=0.35\pm0.05$, for which the solid 
(dashed) curve shows a calculation (Vanderhaeghen \etal 2001) with 
$\xi$-(in)dependent GPDs (Vanderhaeghen \etal 1999). 
The long-dashed curve shows a calculation including twist-3 effects. 
Other curves are for other kinematics. The Figure is taken from 
(Elouadrhiri 2002).}
\vspace*{3mm}
\end{figure}
It aims at a precise check of the 
$Q^2$-dependence of cross section differences
in the reaction $e \, p \rightarrow e \, p \, \gamma$, for
different beam helicities. The second experiment, E01-113 using the
CLAS spectrometer in Hall B 
(Burkert \etal 2001), measures in early 2005 the kinematic dependence of 
the beam-spin asymmetry on $t$, $\phi$, and $x_B$, for several 
fixed $Q^2$-bins. Also cross-section differences will be measured. As 
demonstrated in Figure~12,
these dependences will be measured with a precision that will
allow for a discrimination between certain parameter sets of GPD models.
\begin{figure}[htb]
\label{fig:FutCLAS12GeV}
\vspace*{-10mm}
\includegraphics[width=8cm]{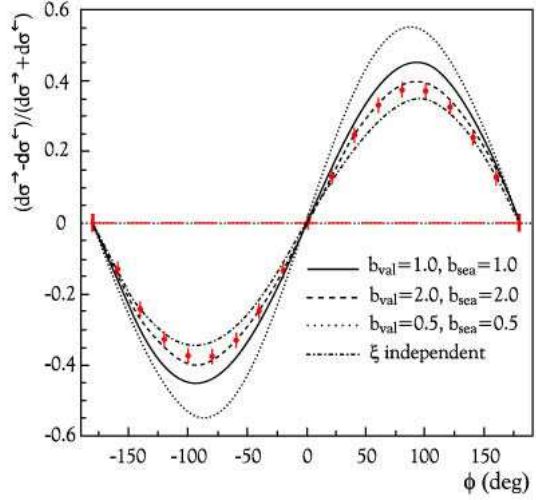}
\centering
\caption{\it {\sc Clas:} Projected statistical accuracy for a 
high-statistics BSA measurement at 11~GeV with an upgraded detector. Bins of
$Q^2=(3\pm0.1)$ GeV$\, ^2$, $W=(2.8\pm0.15)$ GeV, and 
$-t=0.3\pm0.1$ GeV$\, ^2$ are used. GPD calculations (Vanderhaeghen 
\etal 2001) are shown for different combinations of profile 
parameters (Vanderhaeghen \etal 1999). The Figure is taken
from (Mecking 2002).}
\end{figure}

Precision studies of hard exclusive scattering processes at fixed-target 
kinematics are among the main research programs driving the 12~GeV upgrade
of the Continous Electron Beam Accelerator at JLab
(Cardman \etal 2001). The high-duty-cycle and high-intensity beam (electrons
only) will facilitate more accurate measurements of cross sections
and single-spin asymmetries w.r.t. beam helicity and target spin. Running,
{\em eg}, 500 hours with the upgraded CLAS detector at a luminosity
of $10^{35}$ cm$^{-2}$s$^{-1}$ will yield a BSA with a precision about
twice better than that for E01-113 (cf. Figure~13),
so that kinematic dependences can be studied in more detail. Projections
(not shown) demonstrate (Cardman \etal 2001) that using, {\em eg}, 
8 bins in the range $0.2<-t<0.8$ in each of 3x3 cells in the 
($2<Q^2<5$ GeV$^2$,$\; 0.2<x_B<0.6$)-plane, the beam-spin asymmetry may 
be measured with good statistical precision in most of the cells.

No plans are published to also install a positron beam at {\sc JLab}, so that 
no high-precision measurements of beam-charge asymmetries can be expected.

%
%
\subsection{New Results Expected from {\sc Hermes}}
\label{subsec:FutHermes}
%
%
%
Between 2002 and the middle of 2005, {\sc Hermes} data are taken with a
transversely polarised hydrogen target, allowing the evaluation of {\em 
transverse target-spin asymmetries} 
(cf. Equations~\ref{eq:TTSA_N},\ref{eq:TTSA_S}).
Based on an anticipated data sample of about $0.15$ fb$^{-1}$, a first attempt
was made to evaluate the sensitivity to GPDs, in DVCS and hard exclusive
$\rho^0$-production 
on the proton (Ellinghaus \etal 2005). Assuming $u$-quark dominance, the
sensitivity to the GPD $E_u$ was studied and, through a model for it, also 
to the total angular momentum $J^u$ (cf. Equation~\ref{eq:JiSumRule}).
For both reactions, the projected total experimental $1\sigma$-uncertainty 
is equivalent to a range of about 0.12 in $J^u$, so that a significant 
result can be expected.
\begin{figure}[htb]
\vspace*{-0.5cm}
\label{fig:FutBCA+BSA_HERMES}
\includegraphics[width=12cm]{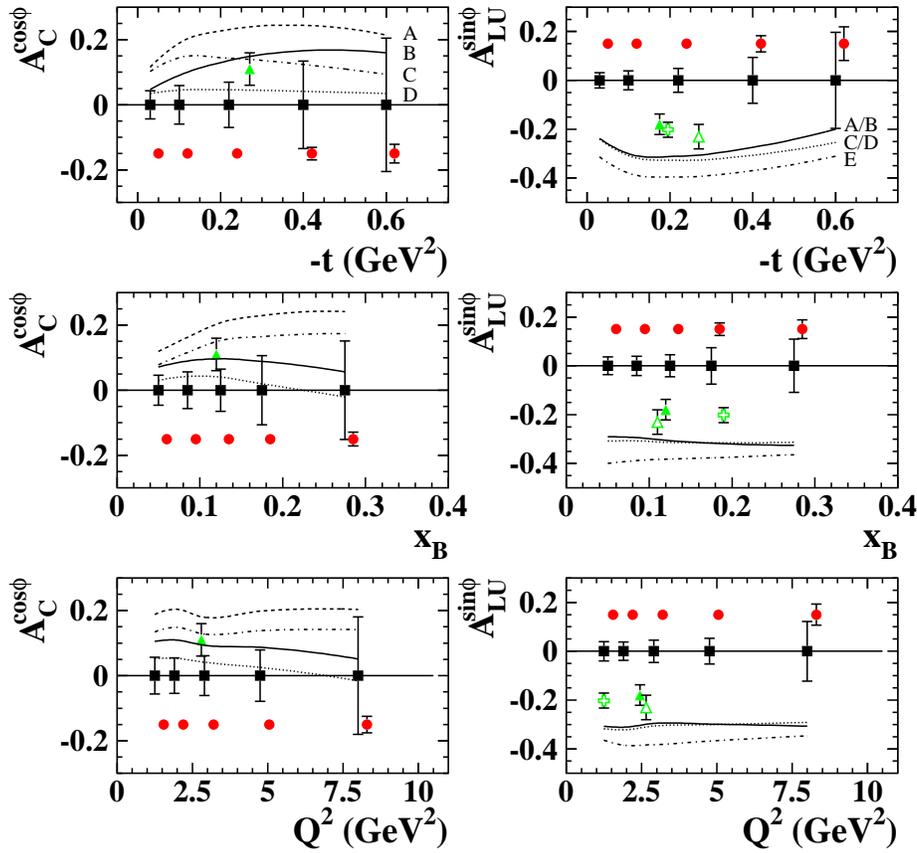}
\centering
\caption{Left (right) panel: Existing data and projections on beam-charge 
(beam-spin) asymmetry on the proton, shown as $\cos \phi$ ($\sin \phi$) 
component. For explanations see text. Notes: i) the more recent 
$t$-dependence shown in Figure~9
is not included here; ii) {\sc Hermes} average kinematics are used for 
the displayed model calculations,
the average {\sc Clas} kinematics are lower (higher) in $Q^2$ ($x_B$) by
about a factor of two, as it can also be seen in the Figure.
The Figure is taken from (Ellinghaus 2004b).}
\end{figure}

The newly built HERMES recoil detector (HERMES Collaboration 2001) will
surround the (unpolarised) internal gas target. By measuring the hitherto 
undetected recoil proton and/or other low-momentum particles, it will 
serve several purposes:

\noindent i) the slow recoil proton can be identified measuring its large 
energy deposition in the (diamond-shaped) double-layer double-sided 
Silicon strip detector.

\noindent ii) in conjunction with a double (stereo)-layer scintillating 
fiber detector
possible additional tracks will be identified, so that the exclusivity of 
the reaction can be established and contaminations from DIS fragmentation
and associated (resonance) production will both be reduced to $\leq$1\%.

For 2005-2007, {\sc Hermes} recoil detector operation is planned with 
an unpolarised hydrogen target, sharing about equally the running time 
between both 
beam charges. In Figure~14
projected  accuracies are shown for 
beam-charge and beam-spin asymmetries, confronted to different 
GPD model predictions that are explained in Table 1. 

\begin{table} [htb]
\begin{center}
\vspace*{2mm}
\small{
\begin{tabular} {|c||c|c|c|c|}
\hline
Model & D--Term & $b_{val}$ & $b_{sea}$ & Ansatz $t$--dependence   \\
\hline
\hline
A & Yes & 1 & $\infty$ & Regge    \\
\hline
B & No & 1 & $\infty$ & Regge  \\
\hline 
C & Yes & 1 & $\infty$ & factorised     \\
\hline
D & No & 1 & $\infty$  & factorised    \\
\hline
E & Yes/No & 1 & 1  & factorised    \\
\hline
\end{tabular}}
\caption{Parameter sets for GPD model predictions calculated 
(Vanderhaeghen \etal 2001) on the basis of 
(Vanderhaeghen \etal 1999) in (Ellinghaus 2004a).}
\vspace*{-4mm}
\end{center}
\end{table}
Error bars shown 
are total experimental uncertainties, {\em ie}, statistical and systematic 
uncertainty added in quadrature. The statistical accuracy at a given point in
one of the variables ($-t,x_B,Q^2$) includes integration over the other two
variables. Existing data already discussed above, 
included for completeness, are shown at average kinematics: Preliminary 
{\sc Hermes} data are represented by closed triangles for BCA 
(Ellinghaus 2002a) and 
2000 BSA (Ellinghaus 2002b), but open triangles for 96/97 BSA 
(Airapetian \etal 2001); open crosses show BSA data from {\sc Clas} 
(Stepanyan \etal 2001). Projected total experimental uncertainties for 
future BCA and BSA results from HERMES are shown in dependence on $-t$, 
$x_B$, and $Q^2$ (Ellinghaus 2004b): Black squares show the
precision of the soon expected final results from 1996-2000 proton data
and red circles show the projected precision for 1 fb$^{-1}$ of data from 
recoil detector running in 2005-2007. Clearly, in the last case a finer 
binning will be possible for lower values of the respective variables.

Somewhat earlier (Korotkov and Nowak 2002a) the azimuthal dependence 
of the beam-charge asymmetry was studied for a certain class of GPD models 
(Vanderhaeghen \etal 1999; Goeke \etal 2001); projections for larger 
values of $x_B$ are shown in Figure~15.
For the models chosen, there seems to be a clear 
sensitivity to the existence of the D-term, although it was also 
shown that the D-term contribution can be
replaced by equivalent tuning of other model parameters (Belitsky \etal 2002).
The right panel of the same figure indicates that from the anticipated data 
set even 2-dimensional dependences can be mapped to some extent, here showing 
the $t$-dependence of the beam-spin asymmetry for two distinct regions in 
$x_B$ (Korotkov 2001). 
\begin{figure}[htb]
\label{fig:FutBCA_H_vsPhi+t}
\begin{minipage}[b]{.45\linewidth}
  \centering \includegraphics[width=\linewidth]
{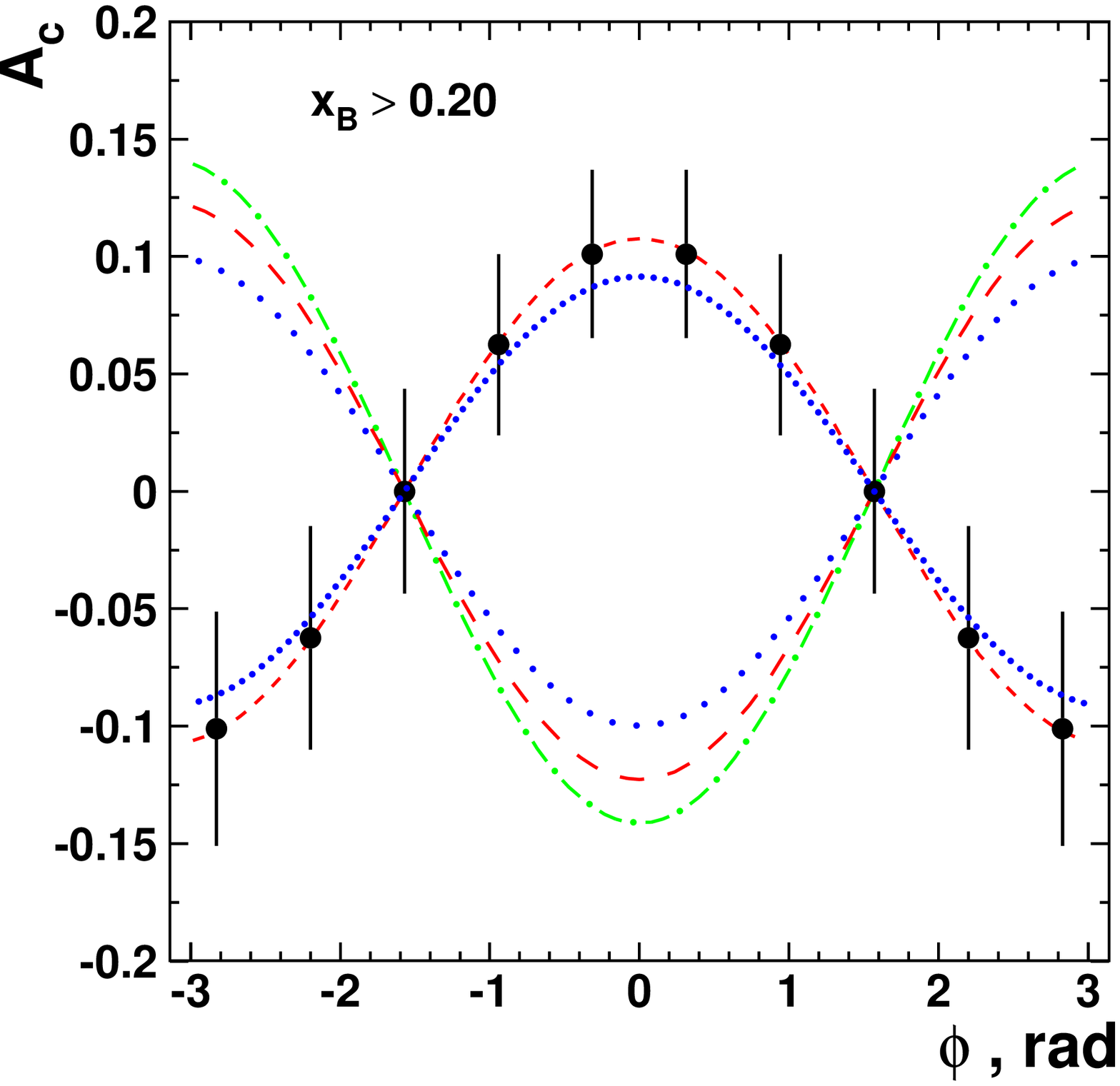}
\end{minipage} \hfill
\begin{minipage}[b]{.45\linewidth}
  \centering \includegraphics[width=\linewidth]
{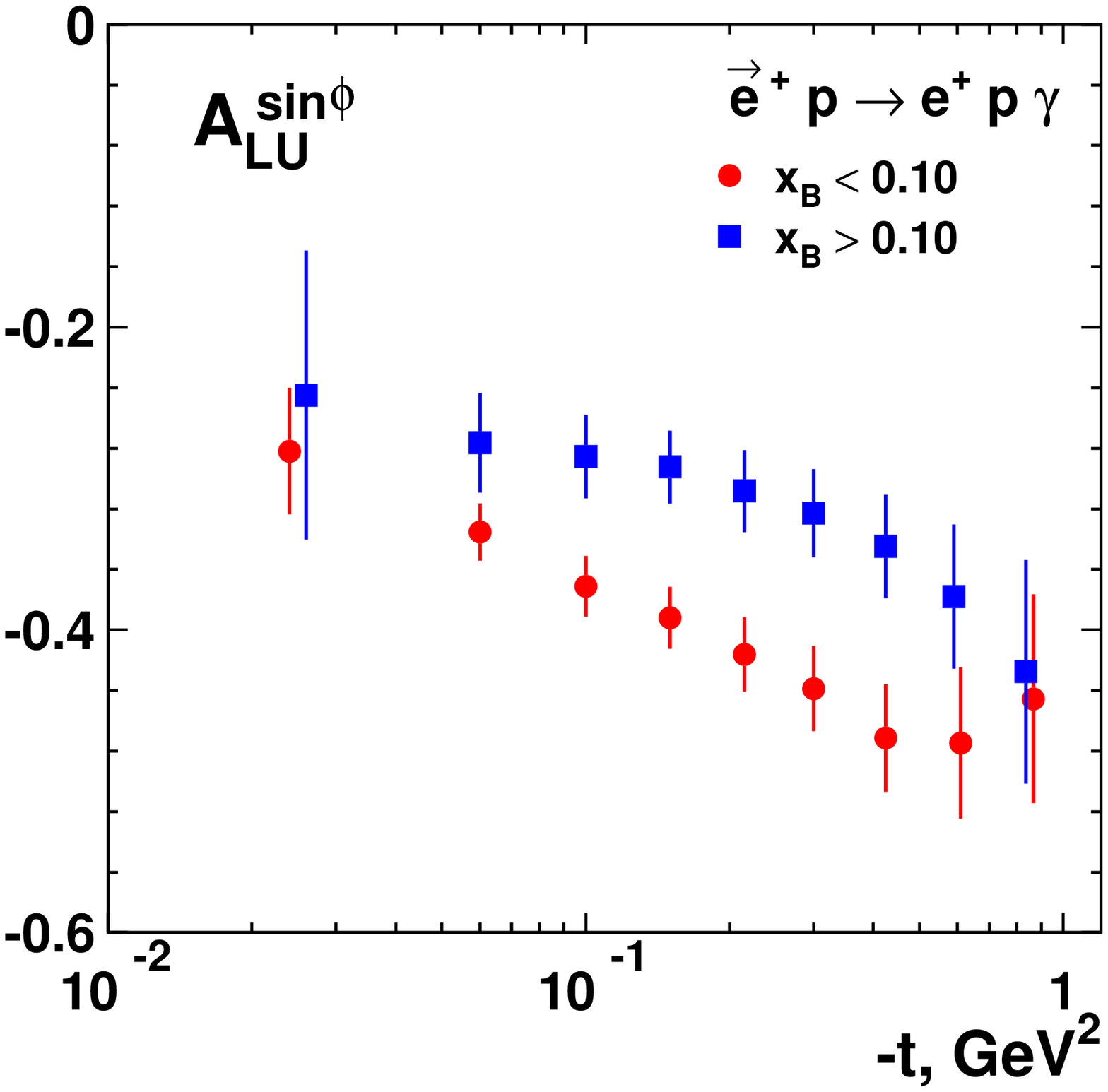}
\end{minipage}
\centering
\caption{\it {\sc Hermes}: Projections for 2005-07 running with a recoil
detector, based on 2~fb$^{-1}$. Left: $\phi$-dependence of beam-charge 
asymmetry in the region $x_B > 0.20$. GPDs calculated
without D-term, where dashed-dotted means $\xi$-independent and 
long-dotted (long-dashed) means $\xi$-dependent with profile 
parameter $b=1$ $(b=3)$, are confronted to those including a D-term, 
denoted by dotted (dashed) instead. Right: $t$-dependence of 
the $\sin \phi$-component of the beam-spin asymmetry, for two distinct
regions in $x_B$. The figures are 
taken from (Korotkov and Nowak 2002a) and (Korotkov 2001).}
\vspace*{-2mm}
\end{figure}
\begin{figure}[htb]
\label{fig:FutImH}
\vspace*{-0.1cm}
\includegraphics[width=6cm]{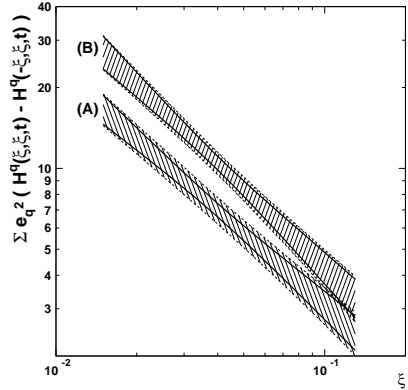}
\centering
\caption{\it {\sc Hermes}: Projected extraction of $Im \, \cal{H}$,
measuring DVCS at {\sc Hermes} with a Recoil Detector in 2005-07, 
based on 2 fb$^{-1}$. The projection B (A) is calculated using a 
$\xi$-(in)dependent GPD, corresponding to the dash-dotted (long-dotted) line 
in the left panel of the previous figure. 
Solid lines enclose the projected 
fully correlated 1$\sigma$ error band in the region $-t \leq 0.15$ GeV$\, ^2$.
The shaded area outside of a band indicates a possible systematic 
uncertainty of the extraction method used.
The figure is taken from (Korotkov and Nowak 2002b).}
\end{figure}

When measuring a beam-spin asymmetry at HERMES kinematics, the imaginary 
part of $\widetilde{M}$ (cf. Equation~\ref{eq:Mtilde} and text thereafter) 
will be dominated by $Im \, \cal{H}$, {\em ie} by
the GPDs $H^q$. This suggests a possible way for a first measurement
of the quantity $\sum_q e_q^2 \, (H^q(\xi,\xi,t)-H^q(-\xi,\xi,t))$. 
Its dependence on the skewedness variable $\xi$ is
shown in Figure~16
(Korot\-kov and Nowak 2002b) for two different GPD parameterisations 
(see caption). Measuring on a proton target, 
$u$-quark dominance can be used to obtain a coarse mapping of the function
($H^u(\xi,\xi,t)-H^u(-\xi,\xi,t)$) as a function of $t$ and $\xi$, {\em ie}
$x_B$. This function is sometimes referred to as `singlet' combination, as 
in the forward limit of vanishing $t$ (and $\xi$) it reduces to the
unpolarised singlet quark PDF $u(x_B)+\bar{u}(x_B)$.

Altogether, the new data set expected from {\sc Hermes} running in 
2005-2007 with a recoil detector will have greatly improved capabilities to 
discriminate between different GPD models.
%
%
%
\subsection{Future DVCS Results from {\sc Compass}}
\label{subsec:FutCompass}
%
%
When considering leptoproduction by muons instead of electrons,
the strength of radiative elastic scattering is reduced by the squared ratio 
of the beam particle masses, $(m_e/m_\mu)^2$ (Mo and Tsai 1969). 
The relative contributions of Bethe-Heitler and DVCS processes to real
photon leptoproduction vary strongly with beam energy, the former 
dominates the latter at electron beam energies of 27.5 GeV 
(Korotkov and Nowak 2002a) and below. Hence at {\sc Hermes} and {\sc Clas}
the DVCS cross section contribution is very hard to access experimentally.
Instead, for a muon beam the DVCS process is already dominant over the 
\begin{figure}[htb]
\includegraphics[width=6cm]{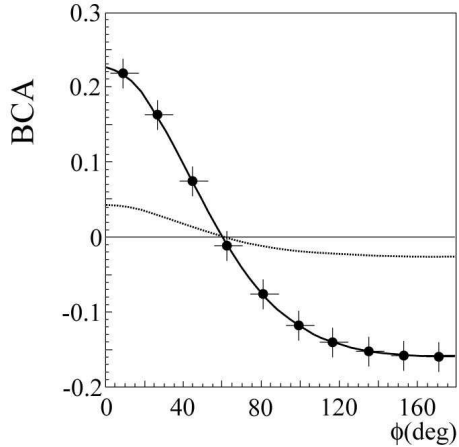}
\centering
\caption{\it {\sc Compass}: Projected statistical accuracy for the
beam-charge asymmetry from 100 GeV running, 3 months each per beam charge, 
with an upgraded apparatus. The statistical uncertainty shown is for the bin
$0.03<x_B<0.07$, $1.5<Q^2<2.5$GeV$\, ^2$ while integrating over 
$0.06<-t<0.3$ GeV$\, ^2$. Solid and dotted curve show a non-factorised and 
a Regge-type GPD ansatz. The Figure is taken from  
(d'Hose \etal 2002).}
\label{fig:FutCompassBCA}
\end{figure}
BH one at an energy of 200 GeV, making {\sc Compass} at {\sc Cern} the 
only set-up that 
is able to measure $|\tau^{}_{DVCS}|^2$ at moderate values of $x_B$. 
At 100 GeV, DVCS and BH contribution are of comparable size, suggesting 
this lower energy for a measurement of the beam-charge asymmetry. Like
in the case of the {\sc Hera} electron or positron beam, the {\sc Cern} 
SPS  muon beam can be produced with either charge. Unlike the former
case, its helicity is fixed and hence always non-zero for individual beam 
charges. The resulting non-zero $\sin \phi$-component in the beam-charge 
asymmetry drops out when symmetrising the BCA, {\em ie,} when calculating 
it only over a range of $\pi$.

A possible GPD experiment at {\sc Compass} (Burtin \etal 2003) would 
use a 2.5 meter long liquid hydrogen target to achieve a luminosity 
comparable to that of {\sc Hermes}, {\em ie,} approximately 
10$^{32}$ s$^{-1}$cm$^{-2}$. Several detector upgrades are necessary to 
reduce photon background from $\pi^0$ and to detect the recoil proton.
The anticipated accuracy of a DVCS cross section measurement at 
$E_\mu=190$ GeV amounts to a few \% (Burtin \etal 2003). Running at
100 GeV will allow to measure the azimuthal dependence
of the beam-charge asymmetry in bins of $x_B$ and $Q^2$ with good
statistical accuracy, as can be inferred from Figure~17
(d'Hose \etal 2002), with the two curves representing two different 
GPD ans\"atze, see caption.
%
%
%
%
\section{Conclusions}
%
%
Deeply Virtual Compton Scattering appears to be the presently best tool
to pursue the in-depth study of the angular momentum structure of the nucleon.
Interpreting the rich body of present and future data within the theoretical
framework of generalised parton distributions, it can be expected that
severe constraints to different ans\"atze and parameterisations will emerge.

\noindent 
Final analysis results from existing {\sc JLab} and {\sc Hermes} data
sets are expected soon to give a first glimpse on kinematic dependences
of beam-spin and beam-charge ({\sc Hermes} only) asymmetries on proton
and deuteron ({\sc Hermes} only). Final data on proton and deuteron 
longitudinal target-spin asymmetries can be expected from {\sc Hermes}, 
as well as on the $A$-dependence of BSAs measured on several nuclei.

\noindent 
{\sc Hermes} data taking with a transversely polarised hydrogen target
in 2003-2005 is expected to yield information on transverse target-spin 
asymmetries, which may be capable of giving first experimental hints on 
the total angular momentum of the $u$-quark.

\noindent 
{\sc JLab} experiments running in 2004 and 2005 and {\sc Hermes} running with 
a recoil detector in 2005-2007 are expected to deliver quite accurate 
kinematic dependences of asymmetries and cross section differences. 
This data will presumably allow a first look to
the $\xi$-dependence of the unpolarised `valence' $u$-quark GPD.

\noindent 
Independent information is expected from possible {\sc Compass} running
in the last quarter of the decade, in particular beam-charge asymmetry and
DVCS cross section will be measured with good precision at moderate values 
of $x_B$.

\noindent 
The collider experiments {\sc H1} and {\sc Zeus} are measuring DVCS at 
very small values of $x_B$. They have obtained the $Q^2$ and 
$W$-dependence of the DVCS cross section and will attempt to measure its
$t$-dependence, as well. Based on recent upgrades they will attempt to obtain
results on beam-spin and beam-charge asymmetries from data taking 
in 2005-2007.

\noindent 
The extraction of firm information on GPDs from experimental data will 
continue to constitute a complicated task. As can be judged from today, 
a major step in precision towards multi-dimensional
mapping of generalised parton distributions can be made once the 
12~GeV upgrade of the JLab electron beam facility will have become reality.
This data will then allow to perform a global fit based on high-precision 
beam-spin (and target-spin) asymmetries, aiming at a simultaneous 
determination of several accessible GPDs in dependence on $t, x_B,$ and 
$Q^2$. Note that a considerable model-dependence will remain for the 
(internal) $x$-dependence of GPDs, as it can be accessed only via 
beam-charge asymmetry measurements that are not possible at {\sc JLab}. 
Nevertheless, eventually completely new knowledge may become available on 
the 3-dimensional angular momentum structure of the nucleon.
%

%
%
\section*{Acknowledgements}
%
I am deeply indebted to M. Diehl and F. Ellinghaus for many beneficial
discussions and most valuable comments on the manuscript. My thanks go to the
University of Glasgow, namely to G.~Rosner and R.~Kaiser, for the invitation 
to present this exciting subject at the I3HP European Topical Workshop, held 
in early September 2004 in front of the most beautiful golf scenery of 
St.~Andrews.

\section*{References}
\frenchspacing
\begin{small}

%
%

\reference{H1 Coll., Adloff C \etal, 2001, 
           {\textit{Eur. Phys. J.}} \vol C24 517.}

\reference{HERMES Coll., Airapetian \etal, 2001, \prl
           \vol 87 182001.}

\reference{Anikin I V \etal, 2000, \pr \vol D62 071501.} 

\reference{Belitsky A V and M\"uller, 1998, \pl \vol B417 129.}

\reference{Belitsky A V \etal, 2002, \np \vol B629 323.}

\reference{Belitsky A V and M\"uller, 2002, \np \vol A711 118c.}

\reference{Bertin P \etal, 2000, CEBAF experiment E00-110.}

\reference{Bl\"umlein J \etal, 1999, \np \vol B560 283; 2000, 
                                    \np  \vol B581 449.}

\reference{Brodsky S J \etal, 1972, \pr \vol D6 172.}

\reference{Burkardt M, 2000, \pr \vol D62 071503; Erratum ibid. D66 119903.}

\reference{Burkardt M, 2003, {\textit{Int. J. Mod. Phys.}} A18 173.}

\reference{Burkert V \etal, 2001, CEBAF experiment E01-113}

\reference{Burtin E \etal, 2003, \np \vol A721 368c.}

\reference{Cano F and Pire B, 2003, \np \vol A721 789.}

\reference{Cardman L S \etal, 2001, {\textit{White Paper: The Science 
          driving the 12~GeV Upgrade of CEBAF}},
          http://www.jlab.org/12GeV/collaboration.html.} 

\reference{ZEUS Coll., Chekanov S \etal, 2003, \pl
          \vol B573 46.}

\reference{Diehl M \etal, 1997, \pl \vol B411 193.}

\reference{Diehl M, 2002, {\textit{Eur. Phys. J.}} C25 223;
Erratum-ibid., 2003, C31 277.}

\reference{Diehl M, 2003, {\textit{Physics Reports}} \vol 388 41.}

\reference{Diehl M and Vinnikov A V, 2004, arXiv:hep-ph/0412162.}

\reference{Diehl M \etal, 2005, {\textit{Eur. Phys. J.}}  \vol C39 1.}

\reference{Diehl M and Sapeta S, 2005, {\textit{in preparation}}.}

\reference{Dittes F M, 1988, \pl \vol B209 325.}

\reference{Donnachie A and Dosch H G, 2001, \pl \vol B502 74.}

\reference{Ellinghaus F, 2002a, for the HERMES Coll., \np \vol A711 171c.}

\reference{Ellinghaus F \etal, 2002b, for the HERMES 
          Coll., arXiv:hep--ex/0212019, {\textit{AIP Conf. Proc. 675}} 303,
          Eds. Makdisi I Y \etal.}

\reference{Ellinghaus F, 2004a, PhD thesis, Humboldt University Berlin/D,
           {\textit{ DESY-THESIS-2004-005}}.} 

\reference{Ellinghaus F, 2004b, for the HERMES Coll., to be publ. 
          in {\textit{Proc. of DIS'04, Strbske Pleso/SK}}.}

\reference{Ellinghaus F, 2004c, for the H1, Zeus and HERMES Colls.,
          arXiv:hep--ex/0410094, 
          to be publ. in {\textit{Proc. of ICHEP'04, Beijing/China}}.}

\reference{Ellinghaus F \etal, 2005, {\textit{in preparation}}.}

\reference{Elouadrhiri L, 2002, \np \vol A711 154c.}

\reference{Favart L, for the H1 Coll., 2004, 
          {\textit{Eur. Phys. J.}} \vol C33 S509.}

\reference{Freund A, 2000, \pl \vol B472 412.}

\reference{Freund A and McDermott M F, 2002, \pr \vol D65 091901; 
          {\textit{Eur. Phys. J.}} \vol C23 651.}

\reference{Freund A \etal, 2003, \pr \vol D67 036001.}

\reference{Goeke K \etal, 2001, 
          {\textit{Progr. Part. Nucl. Phys.}} \vol 47 401.}

\reference{Goloskokov S V and Kroll P, 2005, arXiv:hep-ph/0501242.}

\reference{Guidal M, 2002, \np \vol A711 139c.}

\reference{Guidal M \etal, 2004, arXiv:hep/ph-0410251.}

\reference{HERMES Coll., 2001, {\textit{A Large Acceptance 
          Recoil Detector for HERMES}}, DESY PRC 01-01.}

\reference{d'Hose N \etal, 2002, \np \vol A711 160c.}

\reference{Ji X, 1997, \prl \vol 78 610; 
                       \pr \vol D55 7114.}

\reference{Ji X and Osborne J, 1998, \pr \vol D58 094018.}

\reference{Korotkov V A, 2001, {\em talk at `267. WE-HERAEUS-Seminar on
               Generalised Parton Distributions'}, Bad Honnef (Germany),
               Nov. 19 - 21, 2001; {\em also in:} HERMES Coll. 2001.}

\reference{Korotkov V A and Nowak W-D, 2002a, {\textit{Eur. Phys. J.}} 
           \vol C23 455.}

\reference{Korotkov V A and Nowak W-D, 2002b, \np \vol A711 175c.}

\reference{Kroll P \etal, 1996, \np \vol A598 435.}

\reference{Mankiewicz L \etal, 1998, \pl \vol B425 186.}

\reference{Mecking B A, 2002, \np \vol A711 330c.}

\reference{Mo L W and Tsai Y S, 1969, {\textit{Rev. Mod. Phys}} \vol 41 205.}

\reference{M\"uller D \etal, 1994, {\textit{Fortschr. Phys.}} 
           \vol 42 101.}

\reference{Nowak W-D, 2003, `The Spin Structure of the Nucleon' 37, 
          Ed.s Steffens E and Shanidze R, Kluwer Academic Publishers.}

\reference{Polyakov M V and Weiss C, 1999, \pr \vol D60 114017.}

\reference{Polyakov M V, 2003, \pl \vol B555 57.} 

\reference{Radyushkin A V, 1996, \pl \vol B380 417;
                                 \pr \vol D56 5524.}

\reference{Radyushkin A V, 1999, \pr \vol D59 014030.}

\reference{Radyushkin A V and Weiss C, 2001, \pr \vol D63 114012.}

\reference{Ralston J P and Pire B, 2002, \pr \vol D66 111501.}

\reference{Smith E S, 2003, for the CLAS Coll., arXiv:nucl-ex/0308032,
          {\textit{AIP Conf. Proc. 698}} 129, Ed. Parsa Z.}

\reference{CLAS Coll., Stepanyan S \etal, 2001, \prl
           \vol 87 182002.}

\reference{Vanderhaeghen M \etal, 2001, 
          {\textit{Computer Codes for DVCS and BH Calculations}}, priv. comm.}

\reference{Vanderhaeghen M \etal, 1999, \pr \vol D60 094017.}

\end{small}
\nonfrenchspacing
\end{document}